\documentclass{sig-alternate}
\usepackage{mathptmx} 

\usepackage{fancyhdr}
\usepackage[normalem]{ulem}
\usepackage[hyphens]{url}
\usepackage[sort,nocompress]{cite}
\usepackage[final]{microtype}
\usepackage{cite}
\usepackage{amsmath,amssymb,amsfonts}
\usepackage{algorithmic}
\usepackage{graphicx}
\usepackage{textcomp}
\usepackage{xcolor}
\usepackage{fancyhdr}
\usepackage[hyphens]{url}
\usepackage{graphicx}
\usepackage{lipsum}
\usepackage{multicol}
\usepackage{listings}
\usepackage{algorithm2e}
\usepackage{xcolor}
\usepackage{tikz}
\usetikzlibrary{patterns}
\usetikzlibrary{decorations.pathreplacing,calc}
\usepackage{multirow}
\usepackage[font=small,labelfont=bf]{caption}
\usepackage{color}

\newcommand{\ignore}[1]{}%

\usepackage[bookmarks=true,breaklinks=true,letterpaper=true,colorlinks,linkcolor=black,citecolor=blue,urlcolor=black]{hyperref}

\title{CARGO : Context Augmented Critical Region Offload for Network-bound datacenter Workloads} 
\numberofauthors{2} 
\author{
\alignauthor
Siddharth Rai\ \thanks{This work was done while the author was a post doctoral research at National University of Singapore}\\
\affaddr{Barcelona Supercomputing Center (BSC)}\\
\email{siddharth.rai@bsc.es}
\alignauthor
Trevor E. Carlson\\
\affaddr{National University of Singapore}\\
\email{tcarlson@comp.nus.edu.sg}
}

\begin{document}
\maketitle
\pagestyle{plain}


	\begin{abstract}
		Network bound applications, like a database server executing OLTP queries or a caching server storing objects for a dynamic web applications, are essential 
		services that consumers and businesses use daily. These services run on a large datacenters and are required to meet predefined Service Level Objectives (SLO),
		or latency targets,
		with high probability. Thus, efficient datacenter
		applications
		should optimize their execution in terms of power and performance. However, to support large scale data storage, these workloads make heavy use of pointer connected data structures (e.g., hash table, large fanout tree, trie) and exhibit poor instruction and memory level parallelism. Our experiments show that due to long memory access latency, these workloads occupy processor resources (e.g., ROB entries, RS buffers, LS queue entries etc.) for a prolonged period of time that delay the processing of subsequent requests. Delayed execution not only
		increases request processing latency, but also severely effects an application's throughput and power-efficiency. 
		
		
		To overcome this limitation,
		we present CARGO, a novel mechanism to overlap queuing latency and request processing by executing
		select instructions on an application's critical path at the network interface card (NIC) while requests wait for processor resources to become available. Our mechanism dynamically identifies the critical instructions and
		includes the register state needed to compute the long latency memory accesses. This context-augmented critical region is often executed at the NIC well before execution begins at the core, effectively prefetching the data ahead of time. Across a variety of interactive datacenter applications, our proposal improves latency, throughput, and power efficiency by 2.7X, 2.7X, and 1.5X, respectively, while incurring a modest amount storage overhead.
	\end{abstract}

	\section{Introduction}
	
	The datacenter has become a preferred computing platform to host internet-scale computing services, such as email, social network, enterprise and e-commerce application suites, and more recently, cloud based office productivity and graphics application packages (e.g., MS Office 365, Adobe Creative Suite etc.). The massive scale of these platforms, and the low-latency requirement of interactive applications running on them, demand that these platforms are optimized both in terms of power and performance. Moreover, to support a multi-tenant design (e.g., large number of clients concurrently accessing a server) these platforms distribute the processing across multiple hardware and software components.        	
	\begin{figure}[!hbt]
		\fcolorbox{black}{white}{\includegraphics[width=.47\textwidth]{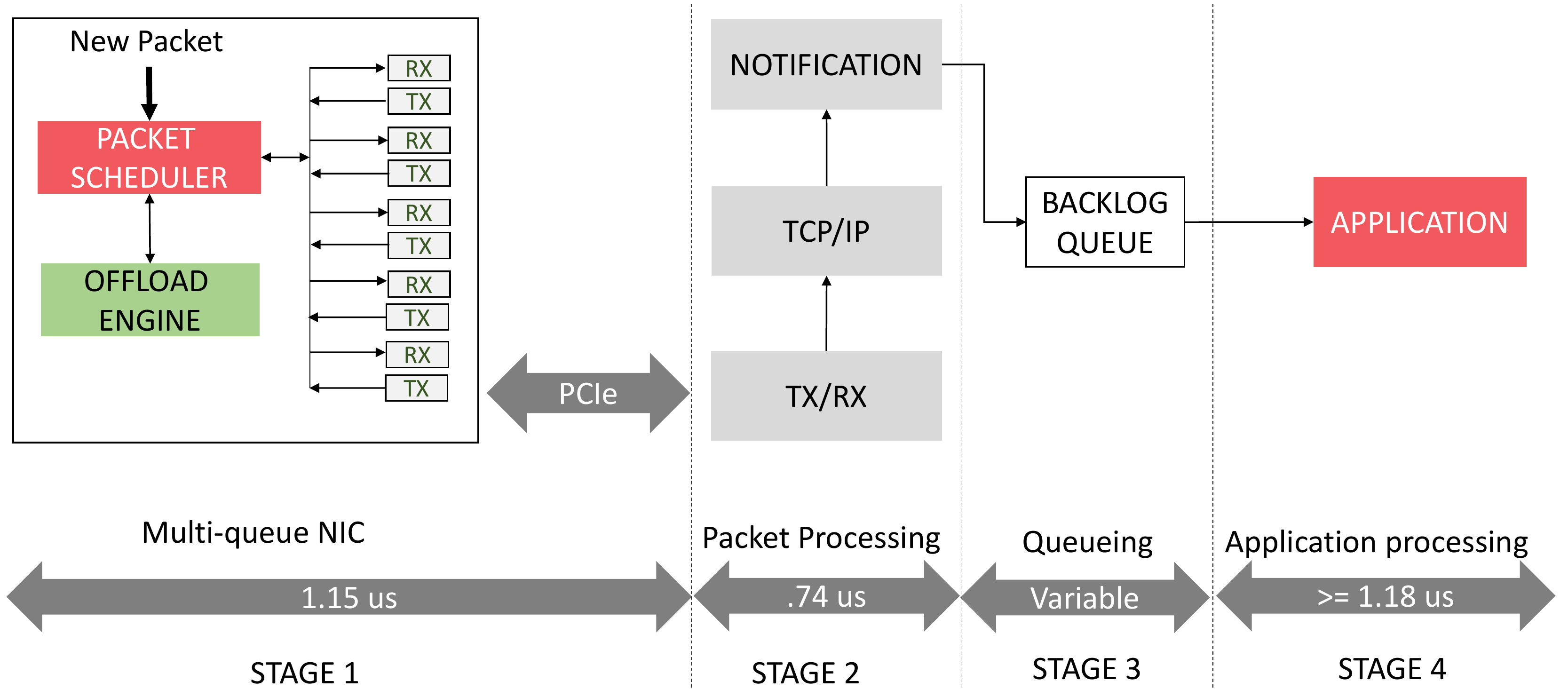}}
		\caption{Stages of a typical request processing pipeline in a datacenter server node. Latency shown in the diagram are obtained using systemtap and conform with~\cite{stackmap}.}
		\vspace{-20pt}
		\label{application-pipeline}
	\end{figure}

	As shown in Figure~\ref{application-pipeline}, a newly received valid packet 
	is enqueued into an appropriate receive ring buffer (Rx) and a CPU is notified about its arrival. Processing at CPU begins by copying the packet from the Rx buffer into an application specific internal buffer\footnote{Data is directly placed in last-level cache of the core, Rx entry only stores a descriptor to in-memory buffer.} (i.e., a user buffer, if zero-copy receive is used; kernel buffer, otherwise.) and performing protocol processing (for a NIC with TCP offload, protocol processing happens at NIC else it happens at the CPU core). 
	From here on, the packet waits for an application thread to become ready and execute the application logic.
	
	
	Key-value stores (e.g., memcached, redis etc.) and in-memory databases (e.g., voltDB, Silo etc.) represent a popular class of latency sensitive datacenter applications used to store dynamic data values required by the contemporary internet-scale services. A key-value store, such as, memcached acts as a front-end caching layer for a large scale distributed database system and thus cuts down the traffic sent to the back-end server. 
	While this application, a key-value store, is designed to efficiently serve infrequently updated data values, others, like an in-memory transactional database, supports the efficient retrieval and modification of frequently changing values. 
	Silo is one such relational database engine supporting transactional operations. Data in Silo is stored in a B+tree structure with a large fanout. 
	Keys in one level of tree acts as prefix for the next level of trees. If the record for a key is found in the same tree, the leaf node stores the corresponding record, otherwise, it points to the next tree level. 
	
	These data-structures allow a wide variety of operations (e.g., from a point query to an expensive range search operation) necessary for efficient data retrieval to support internet-scale services. However, for a large and dynamic dataset, these data-structures require a flexible organization and thus they use a pointer based design. 
	Although using pointers allows for software flexibility, it results in irregular access patterns and can exhibit poor cache locality. As a result, these workloads not only suffer from network processing inefficiencies~\cite{ix, niclatency, zygos} but also lose performance due to poor ILP and MLP characteristics~\cite{thinserversmartpipes, clearingthecloud, meetthewalker} exhibited by the applications. 
	\ignore{	
	To overcome these challenges, recent proposals for datacenter have investigated these platforms for improving power efficiency, reduction in application latency, and workload consolidation. The solutions proposed span from application specific hardware designs to a holistically optimized high-performance platforms.
    }
	
	To this end, 
	we propose Context Augmented Critical Region Offload (CARGO). Our proposal overlaps request wait time and application processing by identifying and executing a program's critical region from the NIC, while, a request waits to execute at the core. Due to this increased concurrency, we are able to improve both application performance and reduce queuing latency. Our proposal obtains the instructions in a program's critical region by dynamically identifying memory, ALU, and control-flow instructions that fall on application's critical path. We further augment these instructions with the register context causing long latency memory accesses. Periodically, instructions, along with register context, are sent to the NIC which executes them for each incoming packet. Since, the major source of degradation in these workloads is irregular memory accesses, executing critical loads early from the NIC timely fills the block into core's private cache and significantly improves L1 cache hit-rate and DRAM bandwidth utilization. Evaluation across a variety of interactive datacenter applications shows that, on average, our proposal is able to improve latency, throughput, and power efficiency of these applications by 2.7X, 2.7X, 1.58X, respectively. 
	
	The rest of the paper is organized as follows. Section~\ref{related-work} presents a background on datacenter architecture and discusses some of earlier work in this area. A motivational study  arguing that a high-performance processor coupled with a low-power NIC has a potential of achieving much higher efficiency is presented in Section~\ref{motivation}. In Section~\ref{cargo}, we present our proposal. Our simulation infrastructure and evaluation is presented in Section~\ref{environ} and ~\ref{evaluation}, respectively. Finally, Section~\ref{conclusion} concludes the paper. 
	\section{Background and Related Work}
	\label{related-work}
	\noindent
	An efficient datacenter requires computing platforms that are easy to program, deliver high power-efficiency, and meet Service Level Objectives (SLO) with high probability. To achieve these goals, the research community has investigated design alternatives that optimize the system in terms of power and performance.
	In the rest of this section, we review some of these proposals and put our contribution into perspective.
	
	\noindent
	\textbf{Accelerators for Offloading:}
	FPGA accelerators and programmable NICs work
	hand-in-hand with CPUs to take advantage of each other's strengths.
	Such a design 
	allows novel application design strategies, like, application partitioning 
	and offloading 
	to NIC cores. Moreover, an RDMA (Remote Direct Memory Access) capable NIC allows executing memory operations on host memory without involving host CPU~\cite{kvdirect}. However, since RDMA primitives are very specific (i.e., memory copy, atomic increment, decrement, etc.), they cover a small class of applications. On the other hand, an FPGA allows synthesizing a specialized data-parallel design for a large class of applications. However, FPGAs are difficult to program and are not well suited for applications with irregular memory access patterns~\cite{pointeronfpga}. 
	Contrary to an FPGA, a programmable multi-core NIC employs low-power embedded-class processors to process network packets at line-rate~\cite{10gignic}. Moreover, ease of programmability and low power consumption of these cores allow offloading a small part of application logic with minimal overhead~\cite{panic}. Thus, a low-power programmable multi-core NIC coupled with a high-performance CPU offers an alternative to an FPGA based design. Motivated by this design choice, our proposal executes judiciously selected instructions with an augmented register context on energy-efficient NIC cores. 

	\noindent
	\textbf{Energy-efficient Design:} Due to excellent energy-efficiency achieved by small cores for mobile platforms, several authors~\cite{mobilesearch, low-power-vs-energy-efficiency, rdbmsenergy, thinserversmartpipes} investigated energy-efficiency of low-power and high-performance processors for datacenter applications. Their study found 
	that smaller cores are less robust towards changing load demand and thus frequently miss QoS targets. Moreover, for complex applications, such as, RDBMS queries, the highest performing design is the most energy-efficient. 
	Another body of work~\cite{dataflowmemcached, memcachedgpu, megakv} investigated FPGA and GPU based accelerators to target high energy-efficiency. However, FPGAs are not only hard to program but also limited by fixed on-board DRAM capacity and the inability of third-party offload engines to support large number of concurrent connections. Compared to FPGAs, GPUs are easy to program and can deliver high throughput and energy-efficiency. However, they are not well suited for pointer intensive applications and exhibit high memory divergence for network-bound applications~\cite{memcachedgpu}. On the other hand, our proposal only executes selected instructions at NIC cores and tries to alleviate any bottleneck due to pointer intensive memory accesses at the CPU core. 
	
%

	\noindent
	\textbf{Software Optimization:} Significance of software components (e.g., TCP/IP processing, packet scheduling, etc.) 
	on performance has also been studied~\cite{ix, zygos}. These proposals optimize data movement between kernel and user address space for protocol processing and load balance packet processing across multiple cores. Since, these solutions are orthogonal to our proposal, we borrow their design and optimize our system to eliminate any software induced inefficiency.
	
	\noindent
	\textbf{Hardware-software Co-design:} 
	Several studies~\cite{highserverutil, mica, billion} investigated hardware-software co-design solutions.  
	These proposals try to eliminate overheads thorough efficient thread scheduling, pinning policies, user-mode protocol processing, and software optimizations that suit application needs. Li et al.~\cite{billion} and Lim et al.~\cite{mica} proposed a full stack redesign of all software components and demonstrated that with all optimizations in place, a 240 core system with 45 MB shared cache can achieve a 1 billion request per second throughput for memcached. Our proposal builds upon this design and borrows design-decisions that apply to our workload. 
	
	\noindent
	\textbf{Microarchitectural and Compiler Optimizations:}	Hardware and software techniques to efficiently execute complex application logic has also been explored~\cite{meetthewalker, imp, swimp, hashjoinprefetching, memoization, runahead, cre, emc, ima}. These proposals try to eliminate micro-architectural inefficiencies for specific data-structure (i.e., hash-table lookup) or memory access patterns (indirect or indexed pointer access etc.). Execution driven prefetching proposals~\cite{runahead, cre, emc} try to utilize processor stalls cycles by executing independent and dependent loads either from the core or from a specialized processor at the memory controller. The Enhanced Memory Controller~\cite{emc} work, which is closest to this work, not only executes dependent loads from the memory controller but also send values to the registers directly and thus eliminates L1 to L3 cache traffic. However, since, these solutions pick instructions from the ROB, they are limited by instructions window lookahead and don't work very well for non-blocking loads (i.e., software prefetches). Contrary to these solutions, our proposal tries to identify critical loads and all other ALU and control-flow instructions needed to execute them. As a result, our solution is able to overcome limited instruction window size and can issue larger set of loads timely. Moreover, since we overlap queuing latency with the execution of critical instruction, we are able to improve both queuing and application latency more effectively.  
	\section{Discussion and Motivation}
	\label{motivation}	
	\noindent
	In this section, we discuss the overheads associated with the request processing pipeline shown in Figure~\ref{application-pipeline}. We confirm that long application latency leads to higher queuing delay and eventually hurts application's power efficiency.
	Given these results, we
	argue that for memory-intensive applications, high-performance CPUs with an optimized memory hierarchy are an efficient design alternative.	
	
	As Figure~\ref{application-pipeline} shows, network processing and PCIe transfer takes 1.15~$\mu s$ while rest of the time is spent in protocol and application processing. As a result, to execute complex application logic at a high request rate,
	a large number of packets that need to be processed concurrently both at the NIC and the processor.
	\begin{table}[!ht]
		\begin{center}
		\fontsize{9}{10}\selectfont
			\begin{tabular}{|c||r|r|r|}	
				\hline
				\multirow{2}{*}{Bandwidth (Gbps)}&\multicolumn{3}{c|}{Required computation (\# Packets)}\\
				\cline{2-4}		
				& $.6~\mu s$ \ & $1.18~\mu s$ & $2.36~\mu s$\\
				\hline
				\hline
				10 & 81 & 94 & 120 \\
				40 & 325 & 376 & 482 \\
				100 & 812 & 941 & 1205 \\
				200 & 1625 & 1883 & 2410 \\
				400 & 3250 & 3767 & 4821 \\
				\hline
			\end{tabular}
			\caption{Count of packets to be processed to meet the line-rate as bandwidth is scaled from 10 to 400 Gbps.}
			\vspace{-15pt}
			\label{motivaton-nic-core-scaling}
		\end{center}
	\end{table}
	Table~\ref{motivaton-nic-core-scaling} shows the number of packets (with 16 byte payload and 48 byte header) that need to be processed concurrently as network bandwidth scales from 10 to 400 Gbps.
	As expected, with increases in network bandwidth, packet count scales up. Moreover, effect of application latency can't be ignored. 
	An increase of .6~$\mu s$ in latency requires 28\% higher throughput (See column 3 of Table~\ref{motivaton-nic-core-scaling}). Moreover, the overhead further grows to 48\% with an additional increase of 1.2~$\mu s$ in latency. 
	\subsection{Network Queuing Analysis}
\begin{figure}[!ht]
    	\centering
    	\includegraphics[width=.4\textwidth]{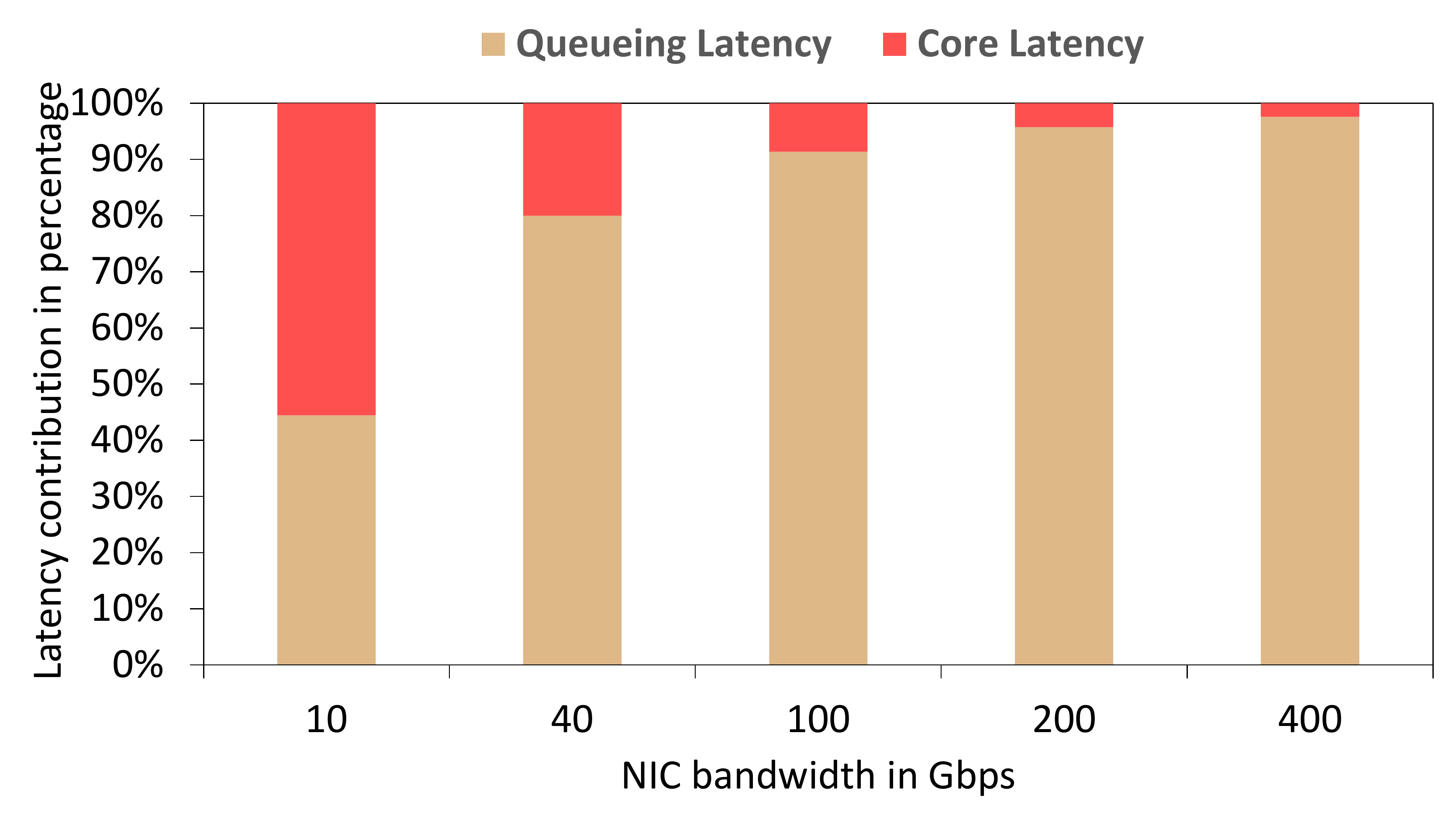}
    	\caption{Contribution of core and queuing latency on a quad-core system as bandwidth is scaled from 10 Gbps to 400 Gbps.}
    	\label{motivaton-latency-component}
    	\vspace{-10pt}
    \end{figure}

	\ignore{
		data-center has become a preferred computing platform to host internet-scale computing services, such as, email, social networking, enterprise and e-commerce application suite, and more recently, cloud based office productivity and graphics application packages(e.g., MS office 365, Adobe creativity suit etc.). Massive scale of these platforms and low-latency requirement of interactive applications running on them demand these platforms to be optimized in terms of power and performance. Moreover, to support a multi-tenant design (e.g., large number of clients concurrently accessing a server) these platforms distribute the processing across multiple hardware and software components.        
		\begin{figure}[!ht]
			\fcolorbox{black}{white}{\includegraphics[width=.47\textwidth]{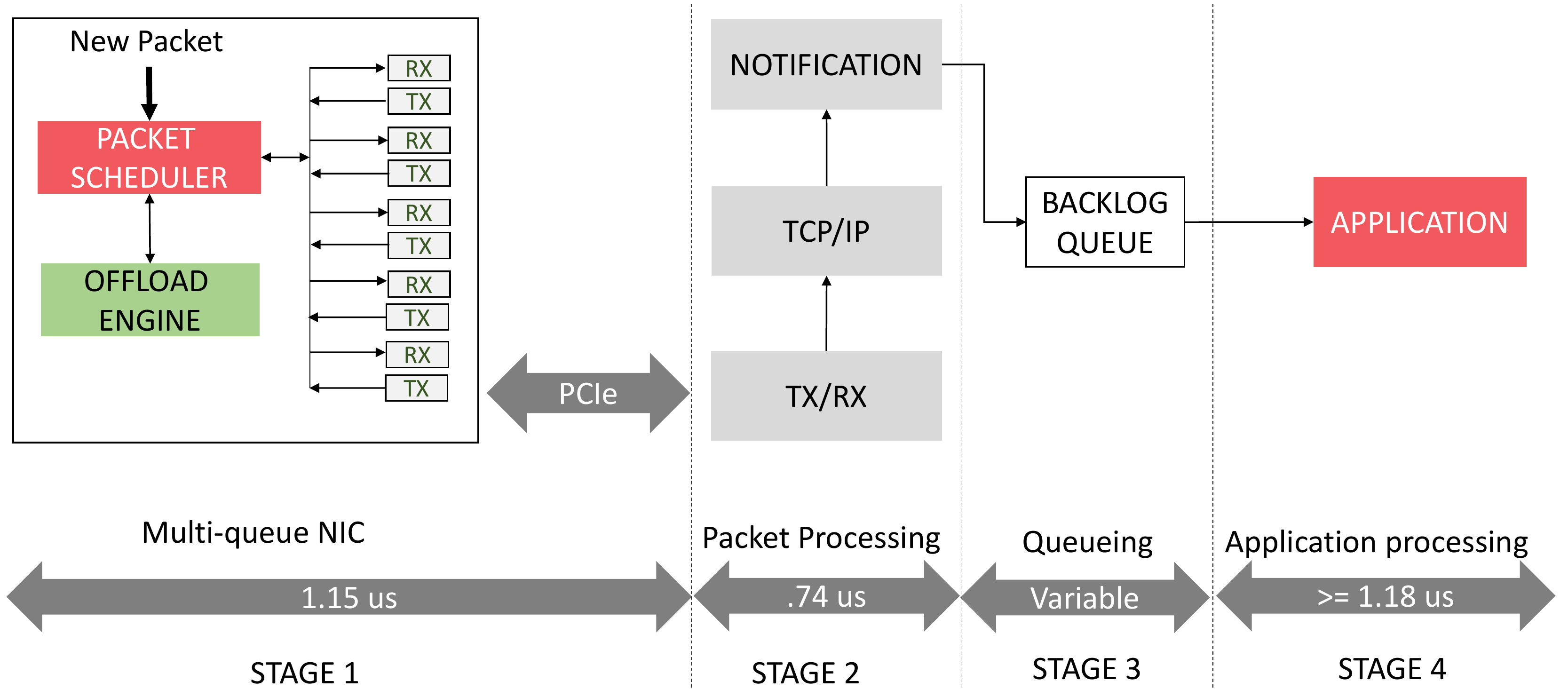}}
			\caption{Stages of a typical request processing pipeline in a datacenter server node. Latency shown in the diagram are obtained using systemtap and conform with~\cite{stackmap}.}
		        \vspace{-15pt}
			\label{application-pipeline}
		\end{figure}
		Figure~\ref{application-pipeline} shows typical stages of a request processing pipeline for a network bound application.
	}%
    \ignore{
	As the figure shows, in the first stage, a newly received valid packet (a packet is validated using checksum computed using an offload routine) is enqueued into an appropriate receive ring buffer (RX) and CPU is notified about its arrival. Processing at CPU begins in the second stage by copying the packet from the RX buffer into an application specific internal buffer (i.e., a user buffer, if zero-copy receive is used; kernel buffer, otherwise. Moreover, data is directly placed in last-level cache of the core, RX buffer only stores a descriptor to the buffer) and performing any protocol related processing (for a NIC with TCP offload, protocol processing happens at NIC else it happens at the CPU core). After protocol processing, packet waits for an application thread to become ready and execute the application logic.
    }
	\ignore{
		As a result, these workloads not only suffer from network interrupt processing inefficiencies~\cite{ix, niclatency, zygos} but also lose performance due to poor ILP and MLP characteristics~\cite{thinserversmartpipes, clearingthecloud, meetthewalker} exhibited by the applications. 
	}
	\ignore{ 
		Nonetheless, improvement in networking technology have gradually enabled high bandwidth and low latency~\cite{lowlatency} support for network interface cards (400 Gbps cards are now available in the market), whereas, due to growing application complexity, processing requests at line-rate has become increasing challenging~\cite{thinserversmartpipes, tailsoftail, stackmap, meetthewalker, panic}. 
	} 
	\noindent
    One way to handle this increasing demand is to spawn more worker threads to process a larger number of packets concurrently. These threads are scheduled on a core as soon as processor resources (ROB entries, RS buffers, etc.) become available. Here we call the time a thread waits for processor resources the queuing latency and the time thread executes on a core as core latency. Figure~\ref{motivaton-latency-component} shows the percentage of time a thread spends queuing and executing at different line-rates. To generate this data, we use an analytical core model proposed for parallel workloads~\cite{threadvscache}. 
    We obtained the model parameters by executing a stock version of memcached~\cite{cloudsuit} on a quad-core processor\footnote{Section~\ref{environ} provides a detailed system configuration.}. Based on the empirical observations, we set the workload specific model parameter $\alpha$ and $\beta$ to 1.5 and 50, respectively.
	\begin{table}[!ht]
		\begin{center}
			\fontsize{9}{10}\selectfont
			\begin{tabular}{|l||r|r|r|r|r|}	
				\hline
				&\multicolumn{5}{c|}{Network Bandwidth (Gbps)}\\
				\cline{2-6}
				& 10 & 40 & 100 & 200 & 400\\
				\hline
				\hline
				Latency ($\mu s$)& 2 &4 &12 &25 & 56\\
				Throughput (MRPS)& 5 &11 &11 &12 & 11\\
				Efficiency (KRPS/W)& 222 &433 &419 &419 & 371\\
				\hline
			\end{tabular}
			\caption{Latency, throughput, and power efficiency achieved at different line-rates.}
			\vspace{-15pt}
			\label{motivaton-core-power}
		\end{center}
	\end{table}
	As the figure shows, across the board, a significant amount of time is spent in queuing delay. Even for the 10 Gbps line-rate, queuing contributes nearly half of the processing latency. Moreover, as more and more packets are processed concurrently, contribution of queuing delay increases sharply. For bandwidth above 100 Gbps, more than 90\% of the latency comes from queuing. Table~\ref{motivaton-core-power} further shows the variation in latency, throughput, and power-efficiency as network bandwidth is scaled up. In spite of the fact that at higher bandwidth more requests are concurrently processed; increased latency saturates throughput and power-efficiency improvement at the 100 Gbps line-rate. These results indicate that queuing latency contributes a sizable amount of the overall packet processing time. Moreover, it increases sharply with small increases in application latency and eventually degrades both throughput and power efficiency. At this point, one might argue that scaling core count might be a better approach, however, increasing the number of physical cores linearly increases core power consumption. Moreover, previous studies~\cite{low-power-vs-energy-efficiency, rdbmsenergy} have found that the highest performing processor also excels in terms of power-efficiency. As a result, for an efficient and scalable datacenter infrastructure, we need to optimize it for both application as well as queuing latency. 
	\subsection{Accelerator Analysis}
	    \begin{figure}[!ht]
		\centering
%
%
     \begin{tikzpicture}[scale=.48]
     \fontsize{6}{6}\selectfont
 \def\nsl{0}
 \def\nst{0}
 \def\nw{1}
 \def\nh{.5}
 
 \def\iohsl{\nsl-.5}
 \def\iohst{\nst-3}
 \def\iohw{3}
 \def\iohh{.5}
 
 \def\llcsl{\iohsl+2.5}
 \def\llcst{\iohst}
 \def\llcw{1}
 \def\llch{.5}
 
 \def\dramsl{\llcsl-.25}
 \def\dramst{\llcst-1.5}
 \def\dramw{1.5}
 \def\dramh{.5}
 
 \def\prsl{\llcsl-.8}
 \def\prst{\llcst+2}
 \def\prw{3}
 \def\prh{3}
 
 \def\mpasl{\llcsl-.25}
 \def\mpast{\llcst+1.5}
 \def\mpaw{.75}
 \def\mpah{.75}
 
 \def\mpbsl{\mpasl+.75}
 \def\mpbst{\mpast}
 \def\mpbw{.75}
 \def\mpbh{.75}
 
 \def\mpcsl{\mpbsl-.75}
 \def\mpcst{\mpbst + .75}
 \def\mpcw{.75}
 \def\mpch{.75}
 
 \def\mpdsl{\mpcsl+.75}
 \def\mpdst{\mpcst}
 \def\mpdw{.75}
 \def\mpdh{.75}
 
 \def\mpsl{\llcsl-.4}
 \def\mpst{\llcst+4}
 \def\mpw{.75}
 \def\mph{.75}
 
 \def\iohtocoret{\nst-2.5}
 \def\iohtocorel{\nsl+.5} 
 
 \def\coretollct{\nst-2}
 \def\coretollcl{\nsl+5.1}
 
 \def\coretollct{\nst-1.5}
 \def\coretollcl{\nsl+2.5}
 
 \def\llctodramt{\nst-3}
 \def\llctodraml{\nsl+2.5}
 
 \draw [black,solid] (\nsl - 1, \nst - 1) rectangle (\nsl+\nw-1,\nst-1+\nh); 
 \draw [black,solid] (\nsl-.5, \nst-.8) node {NIC};
 \node [] (nic) at (\nsl -.5, \nst-.89) {};
 
 \draw [black,solid] (\iohsl - 1.5, \iohst) rectangle (\iohsl+\iohw - 1.5 ,\iohst+\iohh); 
 
 \draw [black,solid] (\iohsl, \iohst + .2) node {PCIe Root};
 \node [] (ioh) at (\iohsl, \iohst+.4) {};
 \node [] (ioh-r) at (\iohsl+\iohw - 1.6, \iohst + .25) {};
 
 \node [] (ioh-ra) at (\iohsl+\iohw-.1, \iohst + .7) {};
 \node [] (ioh-rb) at (\iohsl+\iohw+.7, \iohst + .7) {};
 
 \draw [<->, >=stealth, line width=.5mm] (nic) -- (ioh);
 \draw [black,solid] (\nsl-1.75, \nst - 1.75) node {PCIe x16};
 
 \draw [black,solid] (\llcsl, \llcst) rectangle (\llcsl+\llcw,\llcst+\llch);
 \draw [black,solid] (\llcsl + .5, \llcst+.25) node {LLC};
 \node [] (llc) at (\llcsl + .1, \llcst + .25) {};
 
 \draw [<->, >=stealth, line width=.5mm] (ioh-r) -- (llc);
 
 \draw [black,solid] (\dramsl, \dramst) rectangle (\dramsl+\dramw,\dramst+\dramh);
 \draw [black,solid] (\dramsl + .75, \dramst + .25) node {DRAM};

 \draw [black,solid] (\mpasl, \mpast) rectangle (\mpasl+\mpaw,\mpast+\mpah);
 \draw [black,solid] (\mpasl + .4, \mpast + .4) node {C0};
 \node [] (core) at (\mpasl+.1, \mpast + 1.5) {};
 \node [] (core-l) at (\iohsl+\iohw+.7, \mpast+1.5) {};
 \node [] (core-la) at (\iohsl+\iohw+.6, \mpast+1.6) {};
 
 \draw [black,solid] (\mpbsl, \mpbst) rectangle (\mpbsl+\mpbw,\mpbst+\mpbh);
 \draw [black,solid] (\mpbsl+.4, \mpbst+.4) node {C1};
 
 \draw [black,solid] (\mpcsl, \mpcst) rectangle (\mpcsl+\mpcw,\mpcst+\mpch);
 \draw [black,solid] (\mpcsl+.4, \mpcst+.4) node {C2};
 
 \draw [black,solid] (\mpdsl, \mpdst) rectangle (\mpdsl+\mpdw,\mpdst+\mpdh);
 \draw [black,solid] (\mpdsl+.4, \mpdst+.4) node {C3};
 
 
 
 \draw[<->, >=stealth] (\iohtocorel, \iohtocoret) -- (\iohtocorel, \iohtocoret + 1.25) -- (\iohtocorel + 1.25, \iohtocoret + 1.25);
 
 \draw[<->, >=stealth, line width=.5mm] (\coretollcl, \coretollct) -- (\coretollcl, \coretollct-1);
 \draw[<->, >=stealth, line width=.5mm] (\llctodraml, \llctodramt) -- (\llctodraml, \llctodramt - 1);
 
 \draw [black,solid] (\dramsl, \dramst - .5) node {(A)};
 
 \def\nsl{6}
 \def\nst{0}
 \def\nw{2}
 
 \draw [pattern=north west lines, pattern color=gray] (\nsl - 1.9, \nst - 1) rectangle (\nsl+\nw - 1.5,\nst-1+\nh); 
 \draw [black,solid] (\nsl-.7, \nst-.8) node {NIC + FPGA};
 
 
 \node [] (nic) at (\nsl -.5, \nst-.89) {};
 
 \draw [black,solid] (\iohsl - 1.5, \iohst) rectangle (\iohsl+\iohw - 1.5 ,\iohst+\iohh); 
 
 \draw [black,solid] (\iohsl, \iohst + .2) node {PCIe Root};
 \node [] (ioh) at (\iohsl, \iohst+.4) {};
 \node [] (ioh-r) at (\iohsl+\iohw - 1.6, \iohst + .25) {};
 
 \node [] (ioh-ra) at (\iohsl+\iohw-.1, \iohst + .7) {};
 \node [] (ioh-rb) at (\iohsl+\iohw+.7, \iohst + .7) {};
 
 \draw [<->, >=stealth, line width=.5mm] (nic) -- (ioh);
 
 \draw [black,solid] (\llcsl, \llcst) rectangle (\llcsl+\llcw,\llcst+\llch);
 \draw [black,solid] (\llcsl + .5, \llcst+.25) node {LLC};
 \node [] (llc) at (\llcsl + .1, \llcst + .25) {};
 
 \draw [<->, >=stealth, line width=.5mm] (ioh-r) -- (llc);
 
 \draw [black,solid] (\dramsl, \dramst) rectangle (\dramsl+\dramw,\dramst+\dramh);
 \draw [black,solid] (\dramsl + .75, \dramst + .25) node {DRAM};

 \draw [black,solid] (\mpasl, \mpast) rectangle (\mpasl+\mpaw,\mpast+\mpah);
 \draw [black,solid] (\mpasl + .4, \mpast + .4) node {C0};
 \node [] (core) at (\mpasl+.1, \mpast + 1.5) {};
 \node [] (core-l) at (\iohsl+\iohw+.7, \mpast+1.5) {};
 \node [] (core-la) at (\iohsl+\iohw+.6, \mpast+1.6) {};
 
 \draw [black,solid] (\mpbsl, \mpbst) rectangle (\mpbsl+\mpbw,\mpbst+\mpbh);
 \draw [black,solid] (\mpbsl+.4, \mpbst+.4) node {C1};
 
 \draw [black,solid] (\mpcsl, \mpcst) rectangle (\mpcsl+\mpcw,\mpcst+\mpch);
 \draw [black,solid] (\mpcsl+.4, \mpcst+.4) node {C2};
 
 \draw [black,solid] (\mpdsl, \mpdst) rectangle (\mpdsl+\mpdw,\mpdst+\mpdh);
 \draw [black,solid] (\mpdsl+.4, \mpdst+.4) node {C3};
 
 
 
 \draw[<->, >=stealth] (\iohtocorel, \iohtocoret) -- (\iohtocorel, \iohtocoret + 1.25) -- (\iohtocorel + 1.25, \iohtocoret + 1.25);
 
 \draw[<->, >=stealth, line width=.5mm] (\coretollcl, \coretollct) -- (\coretollcl, \coretollct-1);
 \draw[<->, >=stealth, line width=.5mm] (\llctodraml, \llctodramt) -- (\llctodraml, \llctodramt - 1);
 \draw [black,solid] (\dramsl, \dramst - .5) node {(B)};
 \def\nsl{12}
 \def\nst{0}
 
 \draw [pattern=north west lines, pattern color=gray] (\nsl - 1.9, \nst - 1) rectangle (\nsl+\nw - 1.5,\nst-1+\nh); 
 \draw [black,solid] (\nsl-.7, \nst-.8) node {NIC + FPGA};
 \node [] (nic) at (\nsl -.5, \nst-.89) {};
 
 \draw [black,solid] (\iohsl - 1.5, \iohst) rectangle (\iohsl+\iohw - 1.5 ,\iohst+\iohh); 
 
 \draw [black,solid] (\iohsl, \iohst + .2) node {PCIe Root};
 \node [] (ioh) at (\iohsl, \iohst+.4) {};
 \node [] (ioh-r) at (\iohsl+\iohw - 1.6, \iohst + .25) {};
 
 \node [] (ioh-ra) at (\iohsl+\iohw-.1, \iohst + .7) {};
 \node [] (ioh-rb) at (\iohsl+\iohw+.7, \iohst + .7) {};
 
 \draw [<->, >=stealth, line width=.5mm] (nic) -- (ioh);
 
 \draw [pattern=north west lines, pattern color=gray] (\llcsl, \llcst) rectangle (\llcsl+\llcw,\llcst+\llch);
 \draw [black,solid] (\llcsl + .5, \llcst+.25) node {LLC};
 \node [] (llc) at (\llcsl + .1, \llcst + .25) {};
 
 \draw [<->, >=stealth, line width=.5mm] (ioh-r) -- (llc);
 
 \draw [pattern=north west lines, pattern color=gray] (\dramsl, \dramst) rectangle (\dramsl+\dramw,\dramst+\dramh);
 \draw [black,solid] (\dramsl + .75, \dramst + .25) node {DRAM};

 \draw [black,solid] (\mpasl, \mpast) rectangle (\mpasl+\mpaw,\mpast+\mpah);
 \draw [black,solid] (\mpasl + .4, \mpast + .4) node {C0};
 \node [] (core) at (\mpasl+.1, \mpast + 1.5) {};
 \node [] (core-l) at (\iohsl+\iohw+.7, \mpast+1.5) {};
 \node [] (core-la) at (\iohsl+\iohw+.6, \mpast+1.6) {};
 
 \draw [black,solid] (\mpbsl, \mpbst) rectangle (\mpbsl+\mpbw,\mpbst+\mpbh);
 \draw [black,solid] (\mpbsl+.4, \mpbst+.4) node {C1};
 
 \draw [black,solid] (\mpcsl, \mpcst) rectangle (\mpcsl+\mpcw,\mpcst+\mpch);
 \draw [black,solid] (\mpcsl+.4, \mpcst+.4) node {C2};
 
 \draw [black,solid] (\mpdsl, \mpdst) rectangle (\mpdsl+\mpdw,\mpdst+\mpdh);
 \draw [black,solid] (\mpdsl+.4, \mpdst+.4) node {C3};
 
 
 
 \draw[<->, >=stealth] (\iohtocorel, \iohtocoret) -- (\iohtocorel, \iohtocoret + 1.25) -- (\iohtocorel + 1.25, \iohtocoret + 1.25);
 
 \draw[<->, >=stealth, line width=.5mm] (\coretollcl, \coretollct) -- (\coretollcl, \coretollct-1);
 \draw[<->, >=stealth, line width=.5mm] (\llctodraml, \llctodramt) -- (\llctodraml, \llctodramt - 1);
 \draw [black,solid] (\dramsl, \dramst - .5) node {(C)};
 \end{tikzpicture}
		\vspace{-10pt}
		\caption{Different system configurations with CPU, programmable NIC, FPGA.}
		\vspace{-10pt}
		\label{motivaton-design-space}
	\end{figure}
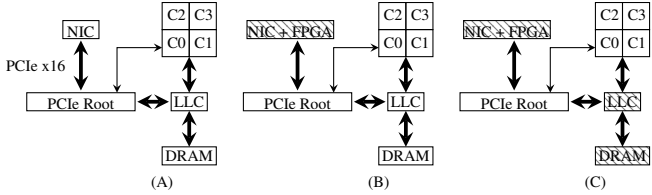
	\begin{figure}
		\centering
		\includegraphics[width=.5\textwidth]{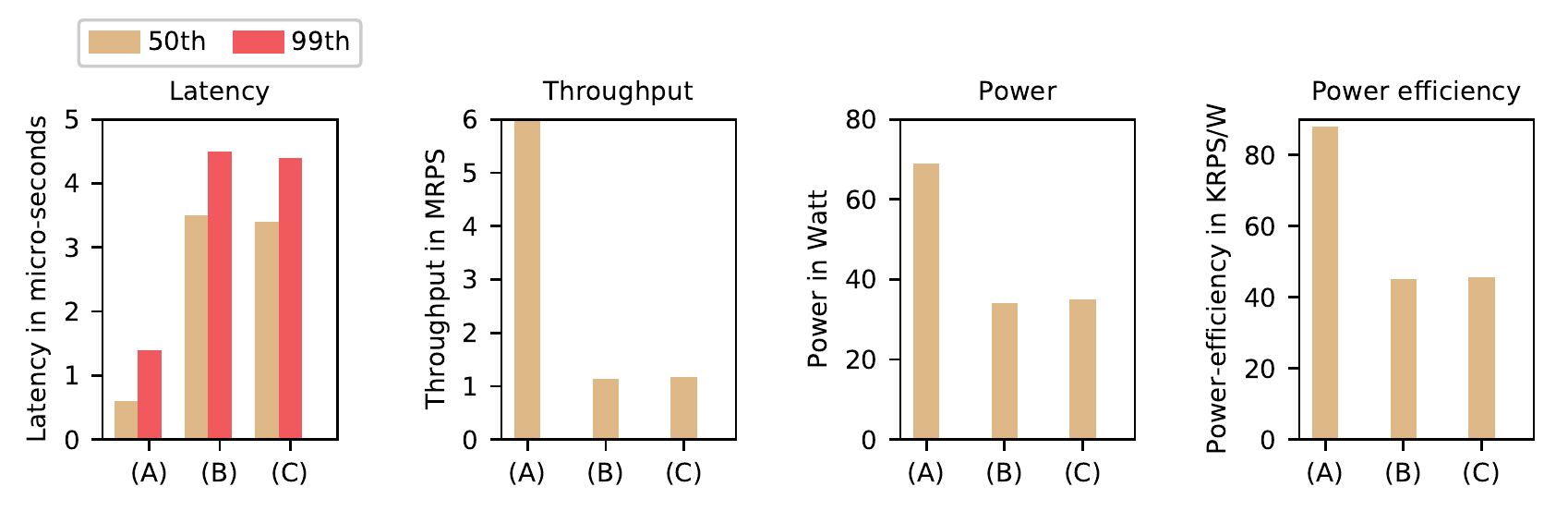}
		\caption{Evaluation of platforms shown in Figure~\ref{motivaton-design-space} using memcached as representative workload.}
		\vspace{-15pt}
		\label{motivaton-design-space-perf}
	\end{figure}
	\noindent
	Since, FPGA based designs are known to achieve superior power-efficiency, we evaluate power consumption and performance of three FPGA and CPU based designs. Figure~\ref{motivaton-design-space} shows the configuration of these designs. The left most configuration (Figure~\ref{motivaton-design-space}(A)) connects a multicore-NIC and a high-performance CPU through a PCIe link. We dedicate one core of the processor to exclusively run the user-mode networking stack. We also enable receive flow steering to maximize application's cache locality. In this design, the NIC is used only for the packet processing and entire application executes on the CPU cores. The second design, shown in~\ref{motivaton-design-space}(B), uses an FPGA-NIC instead of a multi-core NIC and high-performance CPU. The FPGA connects to the CPU through PCIe link and can take advantage of last-level cache (LLC) directly. In this configuration, apart from processing the network packets, the FPGA also executes the application logic (memcached GET operation in our experiments). The third and the final platform shown in ~\ref{motivaton-design-space}(C), again uses a CPU-FPGA hybrid design with an optimized memcached implementation~\cite{fpgahash}.
	To emphasize the difference between designs, we highlight the added or modified component in each successive design (from (A) to (C)) with a different shade. In all designs, the CPU has four cores, clocked at 4 Ghz frequency and features a three-level cache hierarchy (detailed in Section~\ref{environ}). The FPGA runs at 200 MHz and connects to the rest of the system through a PCIe Gen3 x16 interconnect with 250 ns one way latency\cite{pcie}. The DRAM has 180 cycles fixed latency. In all configurations, the NIC has 10 Gbps bandwidth. To able to meet this bandwidth, the FPGA processes 24 packets concurrently\footnote{Based on peak throughput achieved by memcached FPGA appliance~\cite{fpgahash}}. We execute the stock version of memcached~\cite{cloudsuit}\footnote{We chose memcached over silo in this evaluation to leverage previously proposed FPGA memcached appliance and associative hash-tables design (C)~\cite{fpgahash}} for design (A) and (B) and a modified version of memcached with associative hash-bucket for the design (C). In all the experiments we use the Twitter dataset of 4GB as input where caches are warmed-up for 700 million dynamic instructions after which simulation runs in detailed mode for 1 billion dynamic instructions. We quantitatively evaluate each of these designs in terms of latency in micro-seconds, achieved throughput in million request per second (MRPS), power in watt (W), and power efficiency in kilo request per second per watt (KRPS/W).
	
	%
	Figures~\ref{motivaton-design-space-perf} shows the evaluation. We refer to each individual panel in Figure~\ref{motivaton-design-space-perf} using Roman numeral I to IV. 
	Panel I and II present observed latency and throughput, respectively. Power consumption and power-efficiency is presented in panel III and IV, respectively. 
	As figure~\ref{motivaton-design-space-perf}(I) shows, design (A) is the best performing design in terms of latency. It is able to achieve average and $99^{th}$ percentile latency within 2 micro-seconds. Moreover, design (A) is able to achieve more than 3X improvement compared to other designs. In fact, this result is not surprising for a memory intensive workload, such as, memcached. Since the CPU runs at much higher frequency (4 Ghz for CPU vs 200 Mhz for FPGA) and accesses data from L1 and L2 caches at a much lower latency, memcached sees advantages from both. However, we note that even though the CPU clock is 20X faster than the FPGA, achieved improvement in latency is much lower (only 3X). We attribute this behavior to the difference in the execution bandwidth between CPU and FPGA appliance. The FPGA appliance processes 24 packets concurrently by replicating and pipelining the applications logic and meets incoming network bandwidth demand effectively and thus incurs negligible queuing delay. On the other hand, the CPU process 32 packets simultaneously in the timed-interleaved fashion across 32 different threads. As a result, requests at the CPU suffer from longer queuing delay and result in lower than expected performance. We further note that the improvement in latency due to associative hash-table design is marginal (design (C)). On average, latency improves by only 3\% by this optimization. Figure~\ref{motivaton-design-space-perf}(II) presents the resultant throughput achieved by each of these platforms. Due to significantly lower latency achieved by design (A), it also excels in terms of throughput, achieving 6.1 MRPS throughput (4X improvement over FPGA based design). The performance improvement achieved by design (A) is primarily due to higher clock frequency and efficient core caches. However, both a faster clock and SRAM based caches result in significant increase the power and area consumed by the processor. Figure~\ref{motivaton-design-space-perf}(III) shows the power consumed by the evaluated designs. As the figure shows, design (A) consumes 2.5X higher power then all other designs. Nonetheless, design (A) still achieves superior power-efficiency. Figure~\ref{motivaton-design-space-perf}(IV) shows the achieved KRPS/W for each platform. As can be seen in the figure, due to lower application latency, CPU based platform delivers 45\% higher power efficiency than the FPGA. These results suggest that an FPGA based design is not as suitable as other alternatives for memory intensive applications, such as, memcached. Nonetheless, we note that FPGA is able to achieve lowest power consumption than all other designs and thus can't be ignored altogether. To further understand the impact of memory hierarchy on both FPGA and CPU, our subsequent study evaluated design~\ref{motivaton-design-space}(A),(B) with a perfect cache. In this experiment, we made all memory access coming from the FPGA to always hit in the LLC (each access takes into account PCIe and cache access latency) and access coming out of CPU to always hit in DL1 cache. To distinguish these configurations with the original unmodified configurations, we have suffixed all configurations with sequence numbers. We refer to unmodified CPU configuration as CPU:1 and the CPU configuration with a perfect DL1 cache as CPU:2. Similarly, unmodified and modified FPGA configurations are referred to as FPGA:1 and FPGA:2, respectively. Table~\ref{ideal-configs} presents the latency, throughput, and the power efficiency of all configurations.  As the table shows, an ideal LLC for the FPGA results in 50\% reduction in latency, improves the throughput over standard FPGA by 2.4X, and eventually equates CPU:1 configuration in terms of power-efficiency. However, configuration CPU:2 still outperforms all other alternative and improves throughput and energy efficiency over CPU:1 by 4X and 3.5X, respectively. 
	\begin{table}[!ht]
	    \centering
	        \fontsize{9}{10}\selectfont
		\begin{tabular}{|l|c|c|c|c|}
			\hline
			Config & Latency & Thpt & Power & Efficiency\\
			& (ns) & (MRPS) & (WATT) & (KRPS/W)\\
			\hline
			CPU:C1 & 660 & 6.06 & 69 & 87.3 \\
			CPU:C2 & 173 & 23.1 & 83 & 277.6\\
			FPGA:C1 & 2812 & 1.5 & 34 & 43.6 \\
			FPGA:C2 & 1400 & 3.6 & 41.3 & 87.2\\ 
			\hline
		\end{tabular}
		\caption{Evaluation of CPU and FPGA without any memory bottleneck.}
		\label{ideal-configs}
	\end{table}

	In summary, we found that an FPGA is a suitable platform for applications with small working-set or the ones that don't need to access off-FPGA main-memory frequently. However, for the applications with high memory intensity and large working-set, a high-performance CPU with a deep cache hierarchy is a better alternative. Moreover, the difference in achieved performance between CPU:1 and CPU:2 shows that there is a huge performance potential with configuration CPU:1 (3.5X in terms of energy efficiency) that can be tapped through more efficient on-chip cache hierarchy. 
	\begin{figure}[!ht]
	\vspace{-10pt}
		\centering
		\includegraphics[width=.45\textwidth]{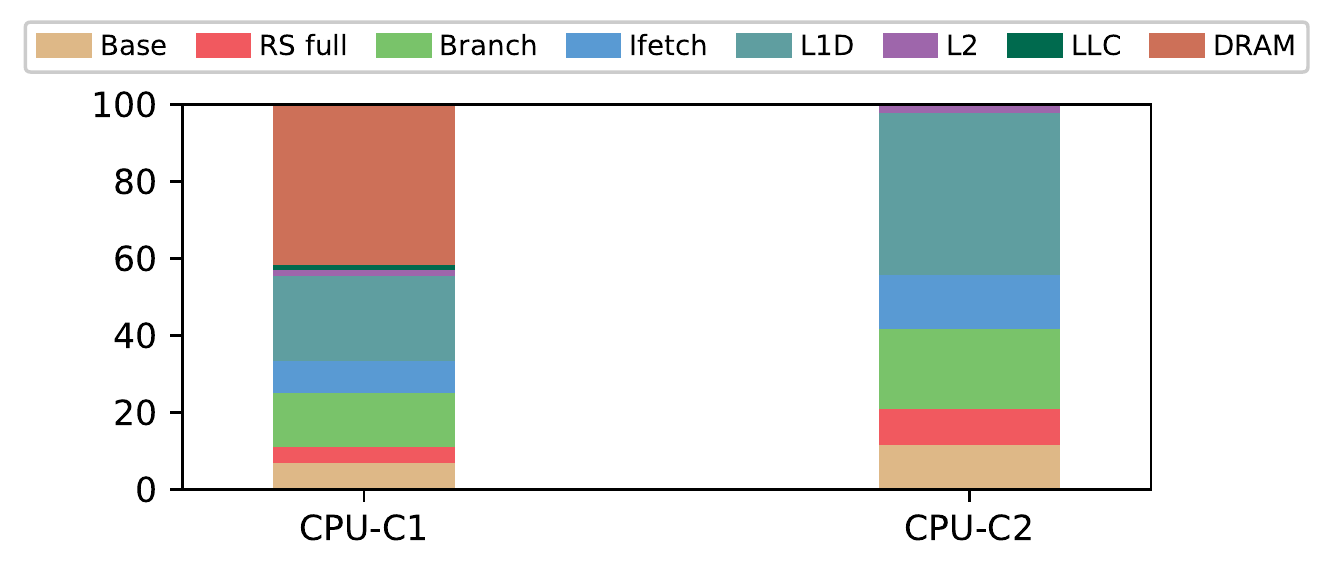}
		\caption{Observed CPI stack for memcached when executed on CPU with configuration 1 and 2.}
		\vspace{-15pt}
		\label{motivaton-cpi-stack}
	\end{figure}
	\subsection{Challenges and Opportunity for Memory Access from NIC}
	\label{memory-bottleneck}
	\noindent
	Figure~\ref{motivaton-cpi-stack} shows the CPI stack corresponding to memcached execution with both CPU:1 and CPU:2 configurations. As the figure shows, for CPU:1, more than 65\% for the cycles are spent in the memory-system alone. On the other hand, if all access from the core can be served by the DL1 cache (CPU:2), memory access overhead drops to 40\%. This reduction not only shortens the critical path length but also removes the bottleneck due to resource contention such as RS buffers and ROB entries. Overall, this reduction improves throughput and energy-efficiency by 4X and 3.5X, respectively.
	\begin{table}[!ht]
		\begin{center}
			\fontsize{9}{10}\selectfont
			\begin{tabular}{|rl|r|l|r|l|}
				\hline
				PC & INST & OPERANDS & TYPE & BB\#\\
				\hline
				{1}& {mov}& {0x364ca0(\%rip),\%rax}&{I-LD}&B0\\
				{2}& {mov}&{    (\%rax,\%rdx,8),\%rbx}&{D-LD}&B0\\
				3& jmp&    SEQ 6& BR&B0\\
				{5}& {mov}&   {0x10(\%rbx),\%rbx}&{D-LD}&B1\\
				6& test&   \%rbx,\%rbx&OP&B2\\
				7& je&     SEQ 20&BR&B2\\
				{8}& {movzbl}& {0x34(\%rbx),\%eax}&{D-LD}&B3\\
				9& cmp&    \%rbp,\%rax&OP&B3\\
				10& jne&    SEQ 5&BR&B3\\ 
				{11}& {movzbl}& {0x2b(\%rbx),\%eax}&{D-LD}&B4\\
				\hline
			\end{tabular}
			\caption{Code region of memcached for hash table lookup.}
			\label{table:memcached-region}
			\vspace{-20pt}
		\end{center}
	\end{table}

	To further understand the nature of memory accesses and their contribution in request latency, table~\ref{table:memcached-region} shows the code snippet of memcached incurring largest number of CPU cycles. For each instruction in the list, its program counter (PC), opcode (OP), and operands (OPNDS) are shown separately. There are five memory access instructions in the list. Instruction 1, is a PC relative load and thus it is accessed on every invocation and can easily be cached. Instruction 2 is an indexed load to register \%rbx depending on \%rdx. Since, \%rdx stores the input dependent hash obtained by applying a function on the input key and results in a random address  and thus is difficult to cache well. The loads at instruction 5, 8, and 11 depend on \%rbx. Register \%rbx either stores a random initial address obtained at instruction 2 or a chain of addresses dereferenced from previous value of \%rbx. As a result, it also leads to sequence of random accesses and often misses in the cache. This input dependent start address and pointer connected design of data-structure make the memory access pattern difficult to track and less amenable to cache. Moreover, the reduced instruction and memory level parallelism (ILP/MLP) delays the processing of subsequent requests and increases queuing latency. 
	The increased queuing latency eventually elongates overall request processing time and degrades throughput.  
	Our proposal named CARGO presented in the following section~\ref{cargo} tries to alleviate this problem through a NIC-core co-design. 

	\section{Context Augmented Critical Region Offload}
	\label{cargo}
	\noindent In this section, we present our Context Augmented Critical Region Offload (CARGO) proposal. 
	\begin{figure}[!ht]
	\vspace{-10pt}
		\centering
		\includegraphics[width=.25\textwidth]{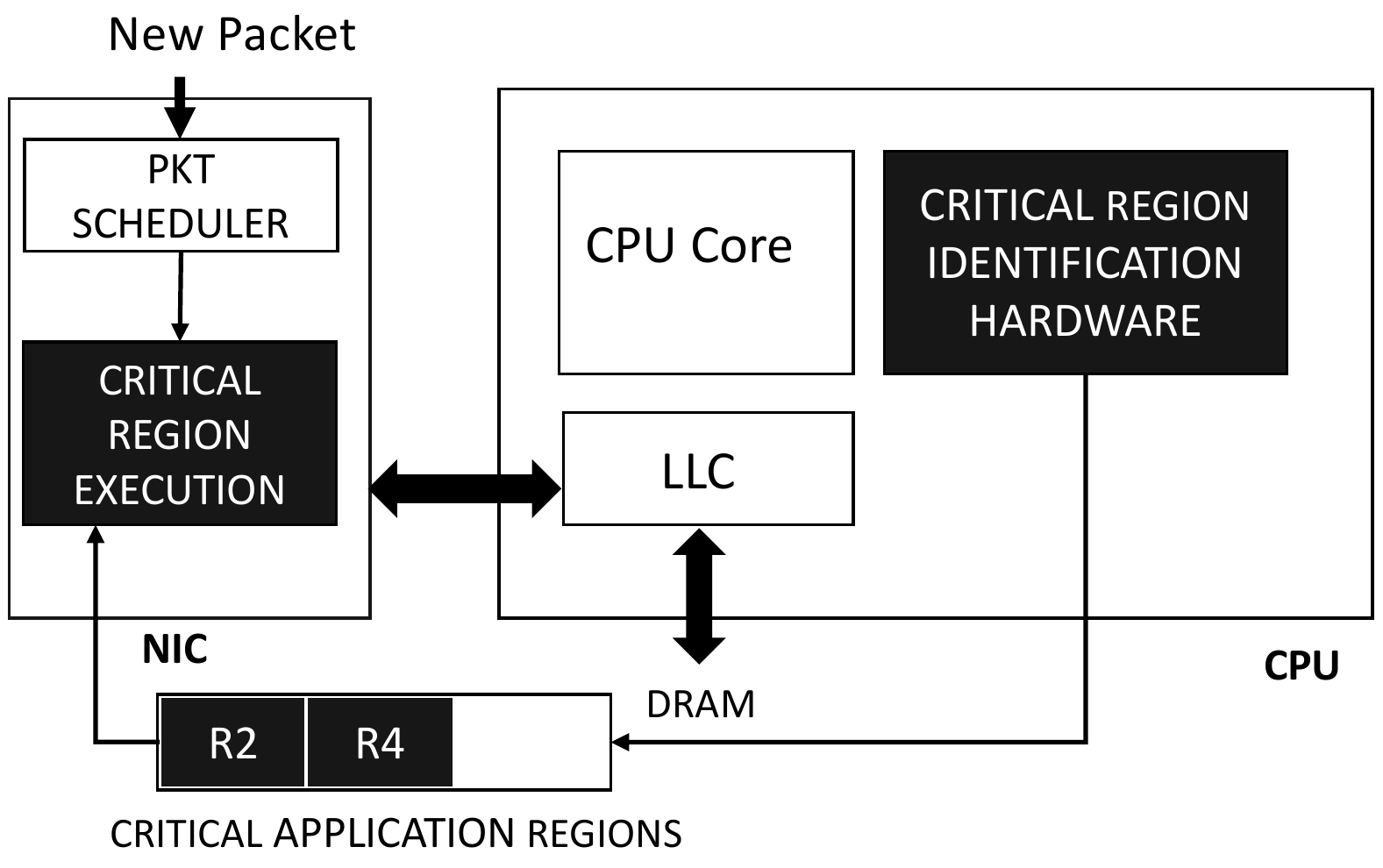}
		\caption{High level view of the proposed design}
		\vspace{-10pt}
		\label{high-level-proposal}
	\end{figure}
	Figure~\ref{high-level-proposal} shows the high-level view of the proposed system. There are two main components in our proposal (shown with black boxes). The first component is located close to the CPU (CRITICAL REGION IDENTIFICATION) and the second component is located on the NIC (CRITICAL REGION EXECUTION). Periodically, identified critical regions 
 (shown as R2, R4) and register state is sent from CPU to NIC for execution. 
	
	The most vital component of our proposal is instruction selection algorithm and the mechanism to obtain register values. To present the insight behind these algorithms, we once again walk through the code region shown in table~\ref{table:memcached-region} contributing highest number of execution cycles with a large fraction coming from memory accesses (refer to CPI stack shown in Figure~\ref{motivaton-cpi-stack}). Semantically, this region maps to the hash-table access part of memcached key lookup routine. In all, there are five different kind of instructions in this region (shown in TYPE column of the table). I-LD instruction corresponds to an independent indirect load, whereas, D-LD instruction depends on a previous instruction. BR is a control transfer instruction. OP and REG are ALU and register to register copy instructions, respectively. There is one instruction of type I-LD and four instruction of type D-LD. Out of four D-LD instructions, instruction 2 depends on instruction 1 and instruction 5, 8, and 11 depend on instruction 2. Rest of the code generates intermediate temporary values and performs logical operation to decide the control flow of the program. 
	\begin{figure}[!ht]
		\centering
		\includegraphics[width=.4\textwidth]{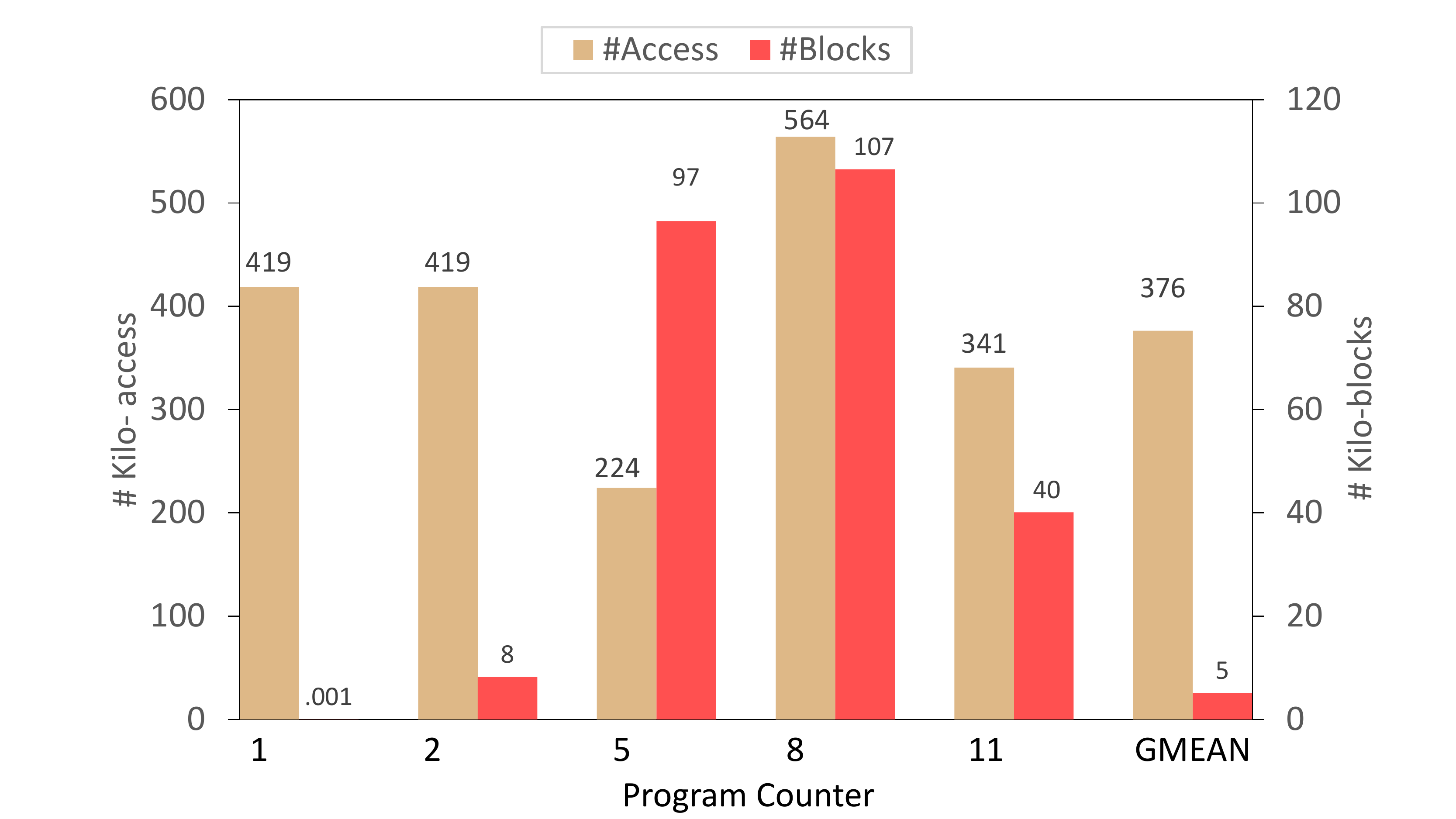}
		\caption{Number of memory access and unique blocks accessed by the sequence of load PCs shown in Table~\ref{table:memcached-region}.}
		\vspace{-5pt}
		\label{per-pc-access}
	\end{figure}
	Figure~\ref{per-pc-access} characterizes load instructions of the region in terms of access count and the number of unique cache blocks sourced by them. Each group of bar on the x-axis presents data corresponding to one load instruction from the list (e.g., bars labelled PC 1 presents data corresponding to instruction with PC 1 in the list and so on and so forth). Left bar in each group correspond to number of accesses, whereas, the right bar correspond to the number of unique cache blocks. The primary and secondary y-axis shows number of access and unique blocks, respectively.  
	
	As the figure shows, number of accesses across PCs vary significantly. On average, each PC sources 376K accesses, with a maximum of 564K (PC 8) and a minimum of 224K (PC 5) accesses. This variation indicates that even though most of the loads are dependent on one particular parent load, their execution is highly control flow dependent and thus readiness of the first instruction of the chain alone cannot determine start of the execution of dependent instructions. Right bar in the figure provides an estimate of the amount of data required to execute these instructions. Interestingly, variation in block count is much larger than that of access count. Blocks count ranges from a maximum of 106K (PC 8) to a minimum of a single block (PC 1, instruction at PC 1 accesses the base address of the hash table, thus it remains static throughout the execution) with an average of 5K across all PCs. The working-set of PCs with fewer unique blocks is smaller and thus can easily be tracked in a small structure completely, whereas, tracking blocks sourced by PCs with large block count would require large storage and should be better left to be generated dynamically (i.e., by calculating address as soon as predecessor load has finished). 
	
	These insights lay foundation of our proposal, which is composed of following three distinct parts: (1) Critical instructions identification, (2) Generating register values for correctly executing these execution, and (3) Executing critical instruction from the NIC.
	\subsection{\textbf{Component 1: Identifying Critical Instructions and Control-flow Information}}
	\label{context-identificatoin}
	\noindent
	Optimizing critical execution path of an applications has been studied in the past~\cite{fields,tune,seng,catch,criticalityvslocality}. 
	They found that long latency loads usually fall on the program's critical path. We leverage this observation for selecting instructions in the critical region. Our algorithm starts by identifying loads that miss often in the L2 cache. Once such loads are identified, other instructions that source values to the instructions currently in the critical region are iteratively added in the list. Finally, we identify control-flow instructions that can alter flow of execution in current critical region and add them to the list. To keep the chain of dependent instructions short, instructions that load values from stack, PC relative address or depend on function argument registers (i.e., registers \%rsi, \%rdi, \%rcx, \%rdx, \%r8, and \%r9, \%rbp, \%rsp) form the start of the chain. This heuristic confines the set of dependent instructions within a procedure boundary and thus keeps it short. 
	\begin{figure}[!ht]
		\centering
		\includegraphics[width=.45\textwidth]{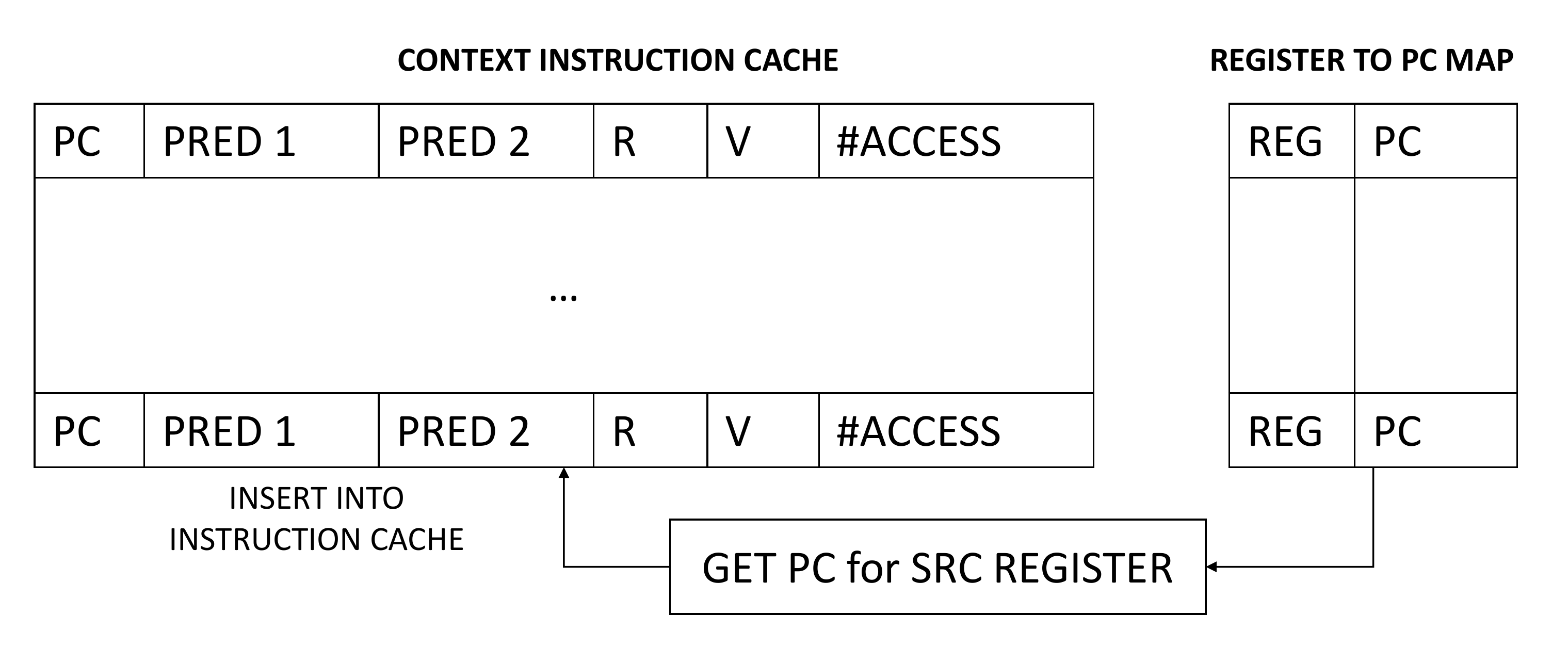}
		\caption{Instruction and data structures used to identify critical code region}
		\vspace{-10pt}
		\label{cargo-structures}
	\end{figure}
	Instructions in the critical region are tracked using a set-associative context instruction cache shown in Figure~\ref{cargo-structures}. The cache is organized as a 256 entry, 16 way set-associative structure. To be able to identify the source of an operand, we maintain a sixteen-entry register to PC map table. Each entry of this table tracks the last PC storing value in a given architectural register. Each entry in the context instruction cache stores the instruction PC, predecessor instructions sourcing operands, ready bit, valid bit, and number of accesses to the entry. 
	An instruction missing L2 cache for data lookup accesses the context instruction cache. If access hits in this cache, block's access count (\#ACCESS field in Figure~\ref{cargo-structures}) is incremented. Otherwise, a new block is allocated in the cache and \#ACCESS is initialized to 1. Predecessor instruction for a newly filled block is obtained by looking up the register to PC map table. In the worst case, predecessor instruction may not be present in the cache. In that case, it is marked for allocation on its next execution. Valid bit is used to indicate occupancy of an entry and referred to while allocating a new instruction. On the other hand, ready bit indicates readiness of an instruction for execution. An instruction with operand sources known and ready is considered ready. 	
	\begin{figure}[!ht]
	\vspace{-10pt}
		\centering
		\includegraphics[width=.2\textwidth]{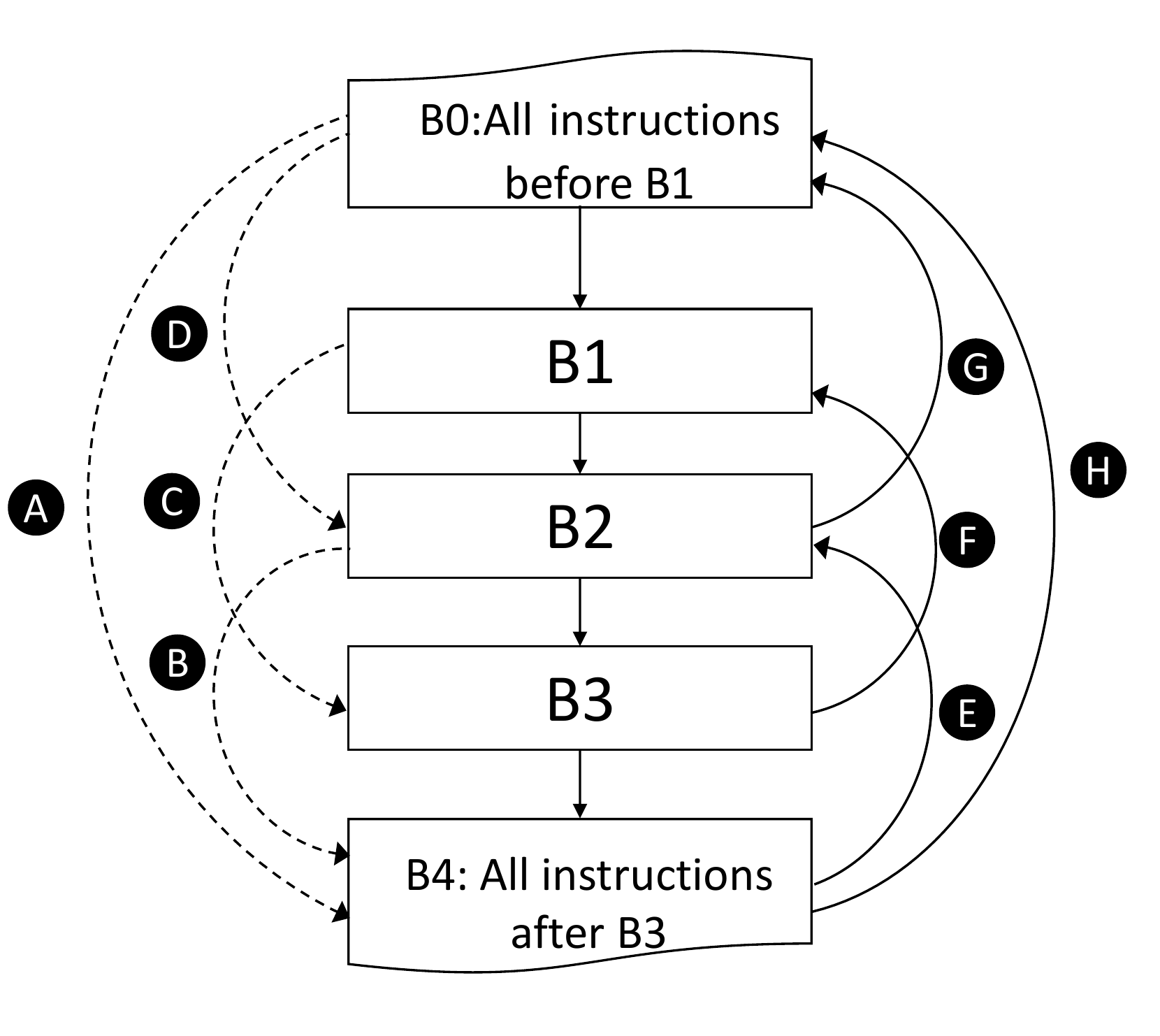}
		\caption{Different control-flow branches need to be handled for the executing instructions in table \ref{table:memcached-region} for any given input.}
		\label{flow-branches}	
		\vspace{-10pt}
	\end{figure}
	
	\noindent
	\textbf{Identifying Control-flow Edges:}
	Once we find the memory and ALU instructions in the critical region, we identify the control-flow (i.e., jmp, jne etc. instructions) that can effect their execution. Since, a branch can originate anywhere in the program and transfer control to the critical region, identifying control-flow is much more challenging than finding memory or ALU instructions. Nevertheless, knowing ALU and memory instruction in advance reduces prospective control-flow paths. 
	Figure~\ref{flow-branches} shows control-flow paths for a sequence of critical instructions in three basic blocks B1, B2, and B3. Block B0 corresponds to all instructions before B1 and the block B4 corresponds to all instructions after B3. Edges A, B, C, and D are the forward edges and edge E, F, G, and H are backward edges. Forward edge A and D originate from an instruction above B1 and have the target in critical region, whereas, edge C has both source and target within the critical region. As a result, if these branches are taken, sequence of instruction execution will change. In contrast to edge B, C, and D, edge A have both source and target outside critical region and thus doesn't effect its execution. Similar to forward edges, back edges that either originate or have target at any instruction in the critical region are relevant. Although, a back-edge like H also bypasses entire region, it is different from edge A and can influence the control flow. Since, edge H transfers control to an instruction before B1 and from there control might eventually reach to B1 again through natural flow of control. On the other hand, such scenario was not possible for edge A because once control passes the basic-block B3, it can never go backwards though a forward-edge. In summary, to correctly identify the control flow for the critical region, we need to track all branches that either have their source or target within the critical region as well as those back-edges that even though bypass but influence its execution.   
	
	\noindent
	\textbf{Tracking Forward Edges:} Our mechanism to track forward-edges is very straight forward. For any control-flow instruction that originates or has a target within the current critical region is also allocated into the context instruction cache. 
	
	\noindent
	\textbf{Tracking Backward Edges:} A backward branch that originate or terminate in current critical region is tracked in the same manner as a forward branch. However, for a branch that bypasses the critical region (i.e., branch H shown in Figure~\ref{flow-branches}), we keep track of register values modified by this branch, so that, when control reaches back to the critical region, we can determine if the used value is supplied by the backward branch or not. Since, at a time only one such backward branch can exist, we checkpoint the register values just before the branch is taken and once the control reaches back to the critical region, the checkpointed register values are compared against the used values. If used values match with the checkpointed values, current backward branch is added to the list of instructions.    
	
	\subsection{\textbf{Component 2: Obtaining Register Values}}
	\label{context-prep} 
        \noindent
	Register values are either received as function arguments or are generated by ALU or memory instructions.
	We follow the standard compiler terminology and call the values received as an argument and generated by the critical instructions as IN and GEN values, respectively. We determine IN values by executing user defined routines (e.g., routines for computing hash values from a key or parsing a query to determine start address of a tree-index) at NIC cores and pass the value to context generation hardware. The hardware compares received values with the IN values seen by CPU during current round of execution. If IN value is found to be coming from a user routines, registers are marked as receiving IN value from the user routine. 
	
\begin{figure*}[!ht]
		\centering
		{\includegraphics[width=.99\textwidth]{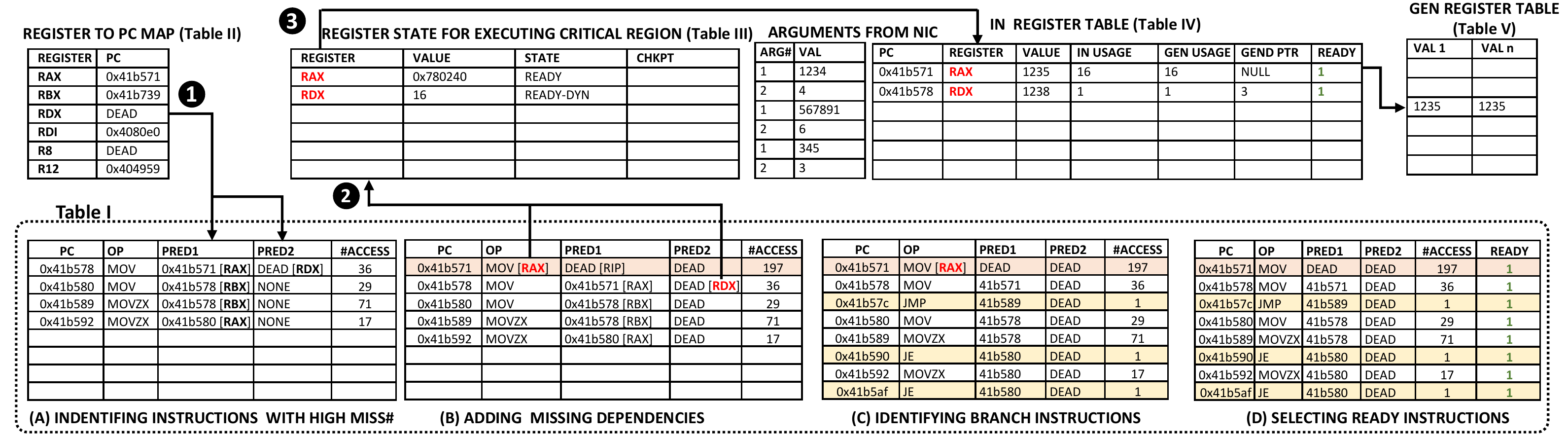}}
		\caption{State of critical cache, register state, and register value predictor used for working example (Subsection~\ref{workingexample}).}
		\vspace{-10pt}
		\label{critical-algo-flow}
	\end{figure*}

	However, for the IN values that can't be determined through user routines (e.g., reference to an object dynamically allocated by the runtime system), we propose a novel register value predictor. 
	Our register value prediction algorithm captures the correlation between IN and GEN values for a given architectural register. An IN value correlated to highest number of GEN values is consider the most probable IN value for that register. 
	There are two table in our design (Table~IV and~V in Figure\ref{critical-algo-flow}). IN values received as arguments are tracked in IN value table, whereas, the values updated during the execution are tracked in the GEN value table. We allow eight different IN values to coexist for a given register. The first IN value for each register is allocated at fixed slot in the table (identified by the register id), whereas, the rest of the values are allocated randomly and connected through the next pointers. When all eight slots for a given register is full, least frequently seen IN value is replaced to make more space. GEN values corresponding to an IN value is tracked in the GEN value table. There can be at most two distinct GEN values for a given IN value. The first entry stores the very first GEN value produced for a given IN value, whereas, the second entry store most recently produced GEN value. We compute IN to GEN transition probability for each slot in IN value table. An IN value with transition probability above 1/8 is considered a valid IN value for a given register.
	Periodically, ready instructions in the cache are identified and critical region is generated. The region along with the register context is sent to the NIC for execution\footnote{Based on empirical observation we set the epoch to 4K L2 misses}. 
	
	\subsection{\textbf{Component 3: Executing Critical Region from NIC}} 
	\label{context-execution}
	\noindent
	The final component of our proposal executes the critical region on NIC cores using the register state received from the critical region identification hardware. 
	On receiving a new packet, a ready context is selected for execution. However, since, state of register may be incomplete (i.e., READY-DYN state) and need to be updated by executing user routines on the newly received packet. We at first execute user routines and update all registers in READY-DYN state and then critical region offload execution is triggered. A critical region offload is executed in a similar fashion as any other NIC offload. 
	We leverage address translation and protection mechanism available at the IOMMU to carry out virtual to physical address translation. Since, NIC is a PCIe device, we utilize PCIe 3.0 Steering Tags (ST) to fill the cache block in to the core's private cache~\cite{steeringpatent}. 
	\subsection{A Working Example}
	\label{workingexample}
	\noindent
	In this subsection, we put together all the components described in subsection~\ref{context-identificatoin}-~\ref{context-execution} and walk through a working example. Figure~\ref{critical-algo-flow} shows the state of context instruction cache (Table~I), register to PC map (Table~II), register state (Table~III), and register value prediction tables (Table~IV-V), as we walk through the example. In the 
	beginning (start of an epoch), Table I has no valid entry; all PCs in Table~II are initialized to INVALID and all other tables are empty. As instructions are executed, Table~II is updated with the most recent PC modifying the register, but with one exception. Every time the smallest PC instruction in Table I is accessed, \%rip, \%rbp, \%rsp, \%dsi, \%rdi, \%rcx, \%rdx, \%r8, \%r9 are initialized to DEAD\footnote{DEAD is used to indicate that register value should be predicted} (these values are later used to identify the root of the dependence chain) and other registers are initialized to INVALID. Instructions with PC 0x41b578, 0x41b580, 0x41b89, and 0x41b592 miss frequently and get allocated in Table~I. For each of these instructions, Table~II is referred and PRED1 and PRED2 fields are updated with current PC~\footnote{For clarity we also show the relevant source and destination registers in square brackets} (shown by arrow~1). For instruction 0x41b578, PC corresponding to \%rdx is DEAD, so, PRED2 is set to DEAD. Whereas, PRED1 refers to PC 0x41b571. Since this PC is not present in Table~I; it is allocated there (shown in Table~I(B)) and its sources are updated later when it is executed again. On every access, instructions in Table~I are checked for readiness. In this case, instruction 0x41b571 is checked first. Since, for this instruction both PRED1 and PRED2 are DEAD, instruction is marked READY. Next, 0x41b178 is checked and also marked READY because PRED1 is DEAD and PRED2 refers to a ready instruction (0x41b571). Similarly, all other instructions in Table~I are checked and their state is updated. When branch instructions (0x41b57c, 0x41b590, and 0x41b5af) are executed, they get added to Table~I, since, their target instructions are already present in Table~I. Moreover, they become READY immediately because their target is ready.  	

	For instructions in Table~I, required register values are tracked in Table~III. There are only two candidate registers (shown in red color in Table~I(B)). First, \%rdx for PC 0x41b578, since it is DEAD but PRED1 is valid. Second, \%rax for PC 0x41b571, since both of PRED1 and PRED2 are DEAD. From here on, these registers are tracked for their values (shown by arrow~2). Table~IV and~V tracks their IN (first value seen) and GEN (subsequent updates) values, respectively (shown by arrow~3). Each newly received in value is compared with the arguments received from the NIC (e.g., hash value of a key). Since IN value of \%rdx matches with the the first argument arguments, its state in Table~III is set to READY-DYN and its value is set to the argument-id (1). For \%rax, whenever a newly received IN value matches with the old IN value, IN usage counter is incremented in Table~IV and for any subsequent GEN values, GEN usage counter is incremented. Since, at this point IN to GEN usage counter exceeds 1/8 \%rax is also set to READY and current IN value is copied to Table~III. At the end of the epoch, all ready instructions with ready register values form one critical region. These set of instructions along with the register values are transferred to the NIC for execution.
	
	On receiving the context from the core, NIC adds the critical region to the list of offloads to be executed and saves the register state into a dedicated area of the on-chip scratch-pad. For each new packet, hash is computed and \%rdx (remember its state is READY-DYN) is initialized. The \%rax is set with the value in register state table (Table~III) received from the core and then the offload is executed. Every time the offload access the off-chip memory, a new PCIe transaction is sent to memory system. The transaction initializes the ST bits of PCIe TLP header to to fill the requested block in appropriate core's L1 cache.
\section{Simulation Environment and Target Workload}
	\label{environ}
%
\begin{table}[h!]
\begin{center}
\caption{Simulation configuration}
\label{tab:environ}
{
\fontsize{9}{10}\selectfont
\begin{tabular}{l|l}\hline\hline
Cores &  4-wide issue/8-wide commit, 256-entry ROB,\\ & 128-entry RS, 4~GHz clock, 4 H/W threads/core\\ \hline\hline
L1 Cache &  32~KB private, 8-way, 2 cycles\\ \hline\hline
L2 Cache &  256~KB unified, 8-way, 3 cycles\\ \hline\hline
LLC  &  1~MB per core, 30 cycles, LRU replacement\\ \hline\hline
NIC & Multi-queue with 6 cores at 166~MHz\\ \hline\hline
PCIe & Gen3~x16 with 250~ns one way latency\\\hline\hline
DRAM: & 45~ns fixed latency, windowed contention\\ \hline\hline 
\end{tabular}
}
\end{center}
\vspace{-15pt}
\end{table}
\noindent
Table~\ref{environ} shows the configuration of our simulation environment. We model a multi-core processor connected with a multi-queue NIC using Sniper Multi-core Simulator~\cite{sniper, snipertaco}. The NIC connects to the CPU over a PCIe Gen3 x16 interconnect. NIC can also access the LLC of the CPU through PCIe root complex. To model the NIC-Core coupled system, we have modified Sniper Multi-core Simulator with appropriate changes. In the modified system, each newly received request is at first processed by the NIC cores and enqueued into an Rx ring buffer. There is one Rx buffer corresponding to each CPU core in the system. From the Rx buffer, requests are dequeued and processed by the corresponding CPU cores. 

Each NIC core has a private 8~KB, 2-way set-associative instruction cache. Control data for packets is stored in an 256~KB 4-way banked scratch-pad memory accessed through a cross-bar. The scratch-pad access latency is two cycles, one cycle to navigate the cross-bar and another cycle to access the memory back. NIC accesses the system memory through a PCIe Gen3 x16 link. We model the PCIe subsystem using a recently published study in this area~\cite{pcie}. To simulate higher NIC bandwidth, we aggregate as many 10~Gbps NICs as are required to meet the target bandwidth.

We evaluate our proposal with a popular in-memory key-value store~(memcached) and an in-memory transactional database system Silo~\cite{silo}. We execute these applications on five different system configurations (ranging from 1 to 16 cores, where each core has 4 hardware threads). Since, a high core configuration can support higher NIC bandwidth, multiple 10~Gbps cards are connected to a high core system. Aggregating multiple NICs allows us modeling higher bandwidth without changing underlying NIC configuration. A system up to four cores connect to one 10~Gbps NIC. For a system with more than four cores, there is one 10~Gbps NIC for each quad core CPU. Thus an eight and sixteen core system supports 20~Gbps and 40~Gbps network bandwidth, respectively. We refer to a configuration by specifying its core and thread count. For example, a configuration with 1 core, 4 thread, and 10Gbps NIC is referred to as 1C/4T. Similarly, a 16 core, 64 thread, and 40Gbp NIC system is referred as 16C/64T and so on and so forth.
We use  memcached version and the data-set accompanied by CloudSuite\cite{cloudsuit} benchmark suit. Memcached is configured to run with 4~GB server memory with GET to SET ratio is fixed to 8:2. 
We use Silo version provided along with the open-source Silo~\cite{silosrc} distribution. We load the database with 16~GB data-set and execute TPC-C queries provided along with Silo source code distribution. 

The part of the application that loads the database and the part executing the queries is explicitly marked in application traces. Before starting the simulation in detailed mode, we warm-up caches and branch prediction tables with the trace region corresponding to the database loading and then the region corresponding to queries is executed in detailed mode. For memcached, each core executes first 700~M instructions in warm-up mode, and subsequent 1~billion instruction in detailed mode. Whereas, for Silo, first 4~billion instructions are executed in warm-up mode and subsequent 1~billion instruction in detailed mode. For each of the benchmarks, we report the latency of queries (or transactions), achieved throughput in request-per-second(or transaction-per-second), and power-efficiency in requests/watt for each of the benchmarks.

	\section{Evaluation}
	\label{evaluation}
	\begin{figure}[!ht]
	\centering
	\fcolorbox{black}{white}{\includegraphics[width=.45\textwidth]{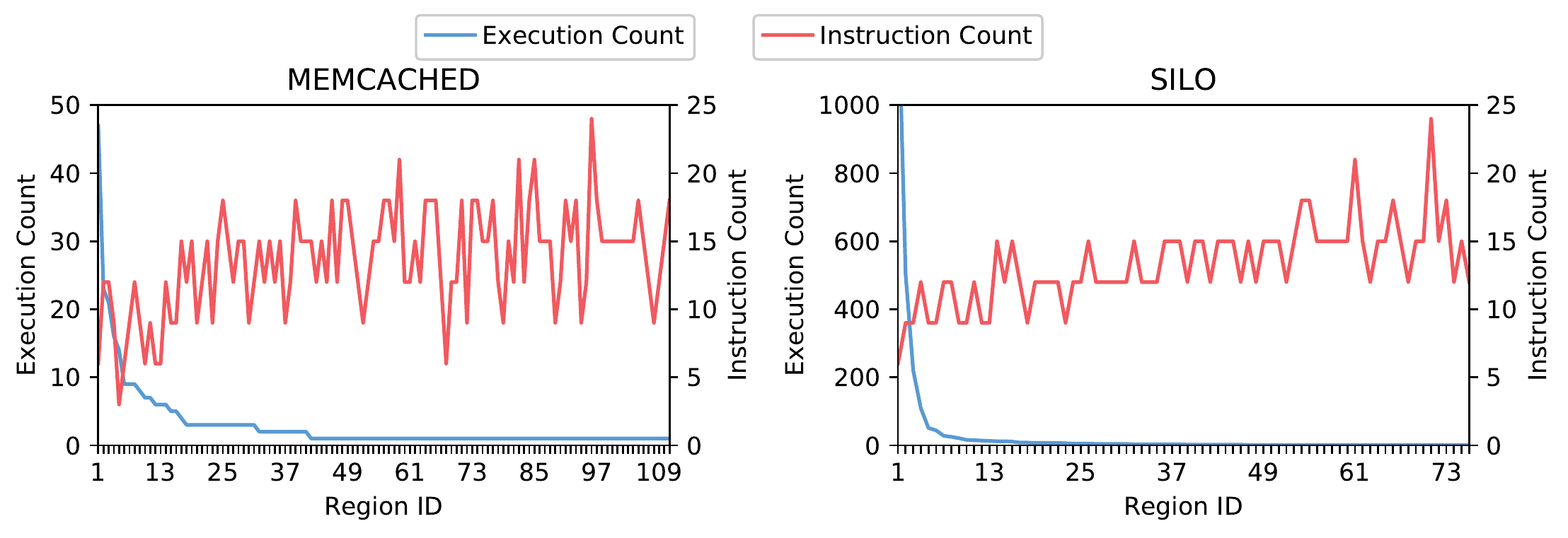}}
	\caption{Characteristics of identified critical regions.}
	\label{critical-regions}
	\vspace{-10pt}
	\end{figure}
	%
	%
	\noindent
	In this section we present the evaluation of our proposal. Since, our design revolves around the identification and execution of critical region, we divide the discussion into two parts. In the first part (Subsection~\ref{critical-region-characteristics}), we present the characteristics of identified critical regions in terms of size, execution count, and constituent instructions. In second part (Subsection \ref{eval-perf}-\ref{overhead}), we present the detailed evaluation in terms of request processing latency, throughput, and power-efficiency improvement and discuss the overheads introduced by CARGO. 	
	\subsection{Characteristics of Critical Regions}	
	\label{critical-region-characteristics}
	\begin{figure}[!ht]
	\centering
	\fcolorbox{black}{white}{\includegraphics[width=.4\textwidth]{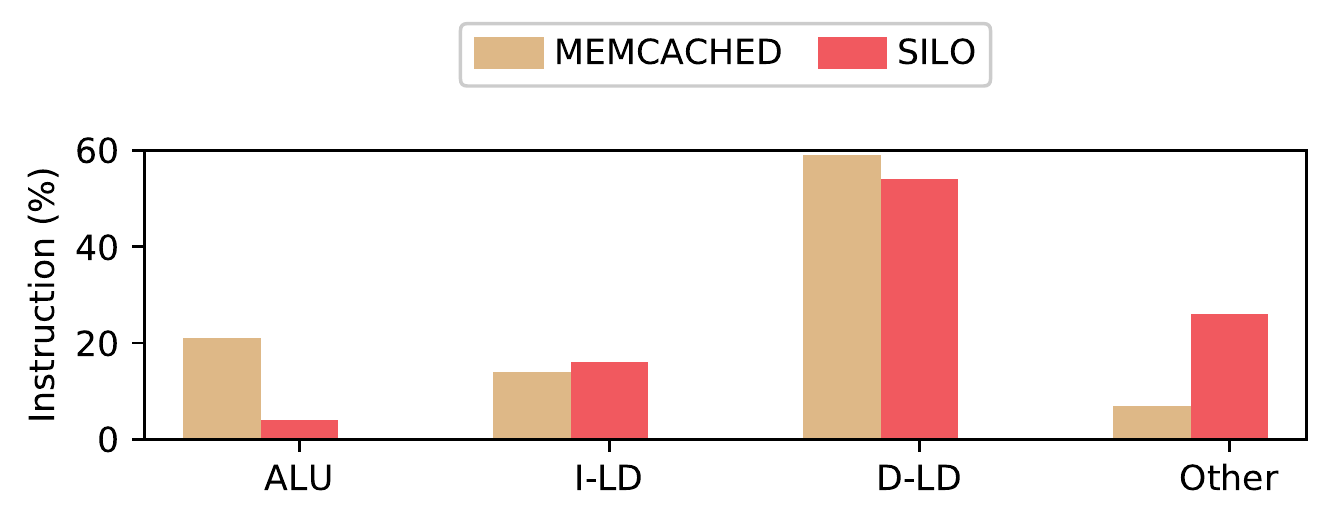}}
	\caption{Instruction classification in critical region.}
	\label{instruction-critical-regions}
	\vspace{-5pt}
	\end{figure}
	\noindent
	Figure~\ref{critical-regions} characterizes the critical regions identified by our algorithm. Each point on the x-axis of the figure correspond to one distinct region. For each region, its execution count and size in terms of instruction count is shown on the primary- and secondary-y axis, respectively. As the figure shows, our algorithm identifies 111 and 76 distinct regions for memcached and Silo (see the maximum region id on the x axis), respectively. Moreover, we note that a small number of critical regions are executed more frequently than others (see the x-axis towards origin). On average, only 2\% and 9\% regions get executed more than 50\% of the time for memcached and Silo, respectively. 
	
	Compared to execution count, region size exhibit little variation. On average, for both, memcached and Silo a region has 13 instructions. Moreover, frequently executed (top 50\%) regions execute only 8 and 6 instructions on average for memcached and Silo, respectively. This data shows that critical regions identified by our algorithm are tiny (6-18 instruction) and a small number of these regions are repeatedly executed.       
        \begin{figure}[!ht]
	\centering
 	\fcolorbox{black}{white}{\includegraphics[width=.4\textwidth]{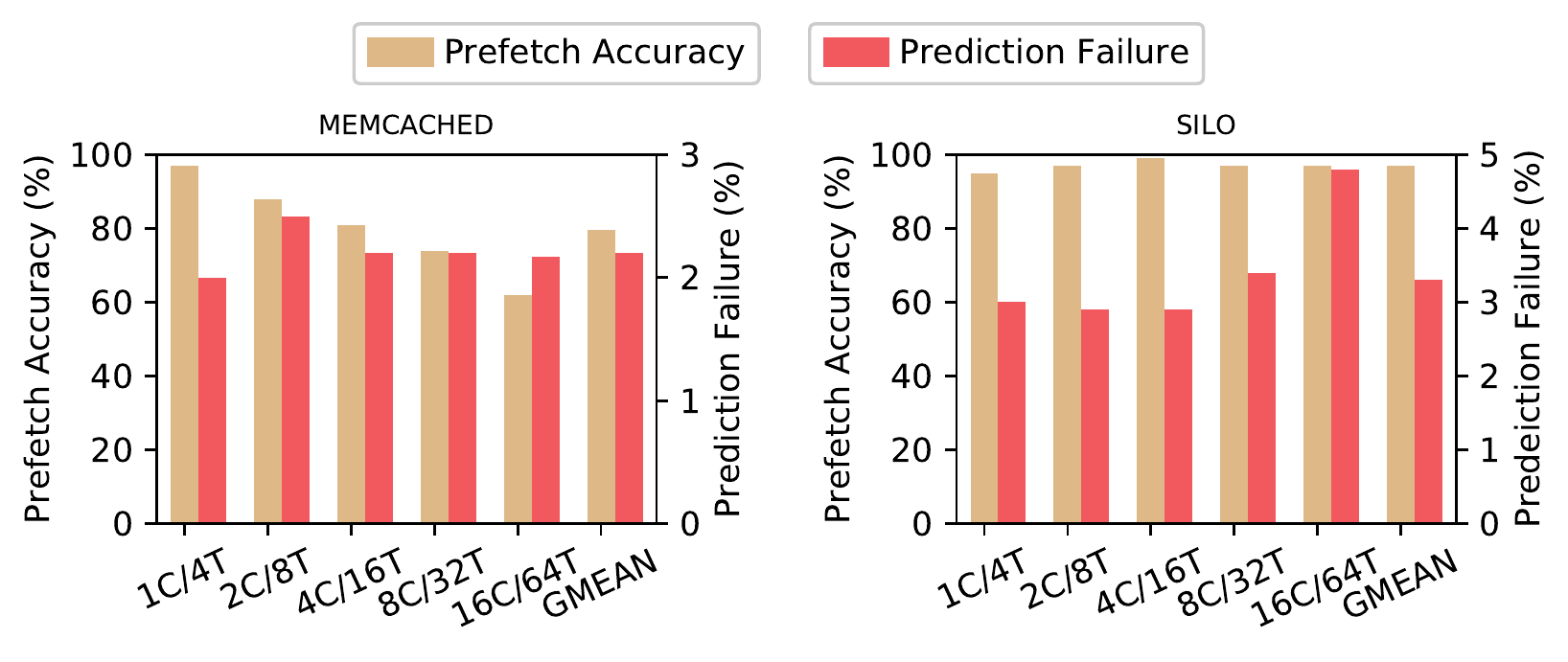}}
	\caption{Observed error in register value prediction}	
	\label{reg-error}
	\vspace{-5pt}
	\end{figure}
	Figure~\ref{instruction-critical-regions} further shows the average distribution of instructions in a critical region. In this analysis, we divide all instructions into four categories, namely, ALU (ALU instructions), I-LD (independent loads), D-LD (dependent loads), and Other (all others clubbed into one category). As the figure shows, instructions are distributed differently across applications. However, combined contribution of I-LD and D-LD instructions is 70\%. Distribution of remaining instructions depend on application. On one hand, ALU instructions make 21\% of the region in memcached, whereas, for Silo only 4\% instructions come from ALU group. These results show that our algorithm successfully identifies both dependent accesses with intermediate ALU instructions and pointer traversals for recursive data structures that involve fewer ALU instructions and more register to register transfer and branch instructions. 
	
	However, correct execution of a critical regions requires correct register values. An incorrect or unknown register value might reduce the coverage and the accuracy of the executed code. Figure~\ref{reg-error} quantifies the error introduced by our register value prediction algorithm. The x-axis of the figure shows evaluated system configurations. For each configuration, primary-y axis shows the percentage of address that are incorrect. Whereas, the secondary-y axis shows the percentage of instructions that fail to execute due to unknown register value. On average, 2.7\% of the instructions failed executing due to unknown register values. Out of the instruction that executed, 7\% of the generated memory accesses are incorrect.  
	\subsection{Performance Improvement with CARGO}	
	\label{eval-perf}
		\begin{figure}[!ht]
		\centering
		\fcolorbox{black}{white}{\includegraphics[width=.48\textwidth]{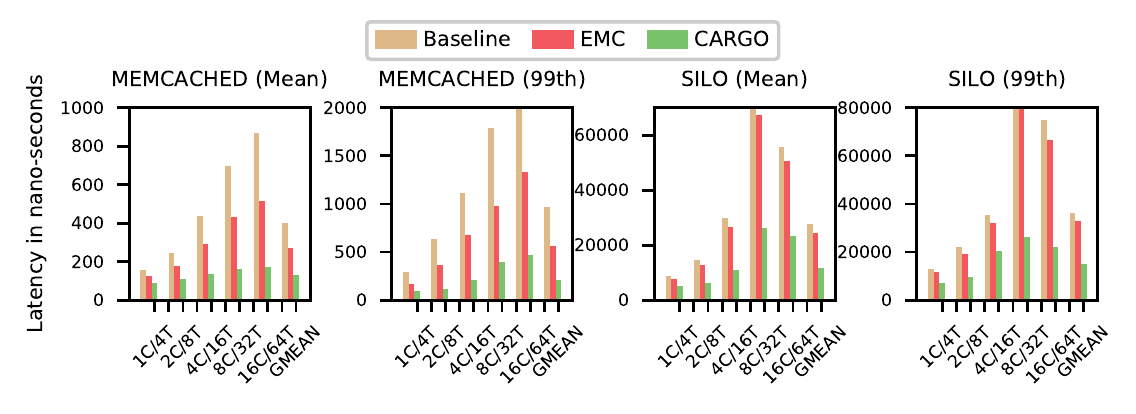}}
		\caption{Observed mean and $99^{th}$ percentile latency for 1 to 16 core configuration.}
		\label{tail-latency}
		\vspace{-15pt}
	\end{figure} 
	\noindent
	Figure~\ref{tail-latency} shows the mean and 99$^{th}$ percentile tail-latency observed by the baseline, EMC, and CARGO. As the figure shows, both mean and $99^{th}$ percentile latency varies significantly across workloads with tail-latency costing nearly 2X higher than the mean. Nevertheless, CARGO is able to improve both tail and mean latencies across all configurations. On average, CARGO reduces mean and $99^{th}$ percentile latency by 2.7X and 3.3X, respectively. 	
	\begin{figure}[!ht]
		\centering
		\fcolorbox{black}{white}{\includegraphics[width=.4\textwidth]{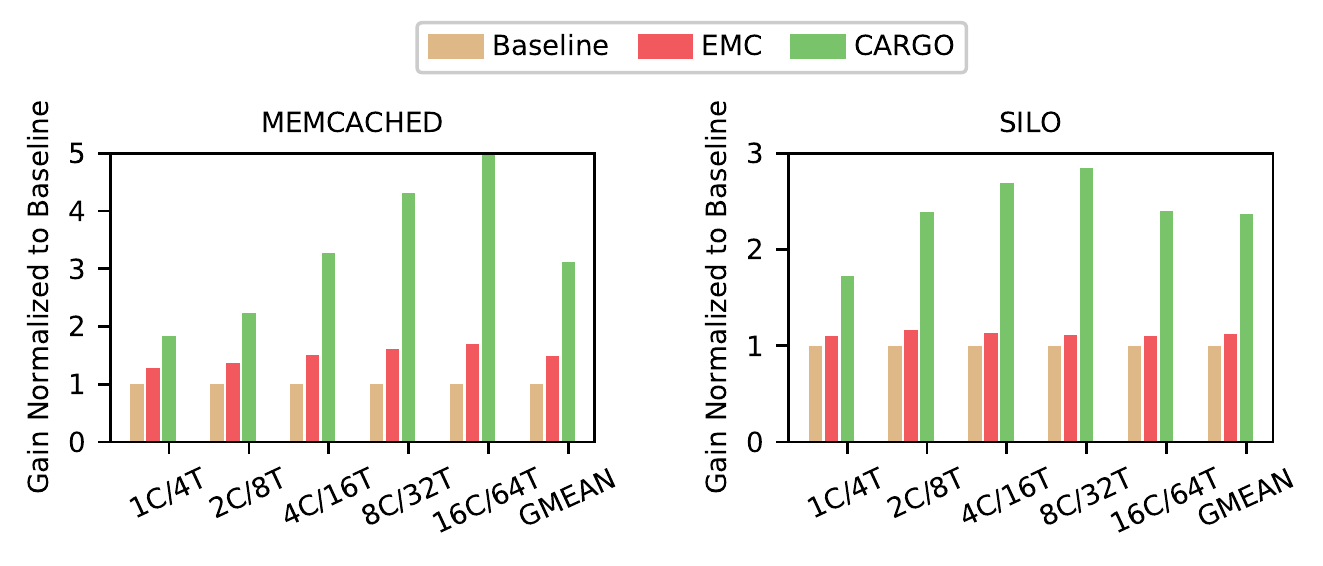}}
		\caption{Throughput normalized to the baseline for 1 to 16 core configuration}
		\label{eval-thp}
	        \vspace{-10pt}
	\end{figure}
	Since, reduced memory access latency eliminates some of the pipeline bottlenecks and improves the number of request processed per second; CARGO is able to achieve significant gain in throughput. Figure~\ref{eval-thp} shows the attained throughput for CARGO and EMC normalized to the baseline. As the figure shows, CARGO achieves 3X and 2.4X improvement in throughput for memcached and Silo, respectively. Whereas, average gain for EMC remains at 42\%. 
	\begin{figure}[!ht]
		\centering
		\fcolorbox{black}{white}{\includegraphics[width=.4\textwidth]{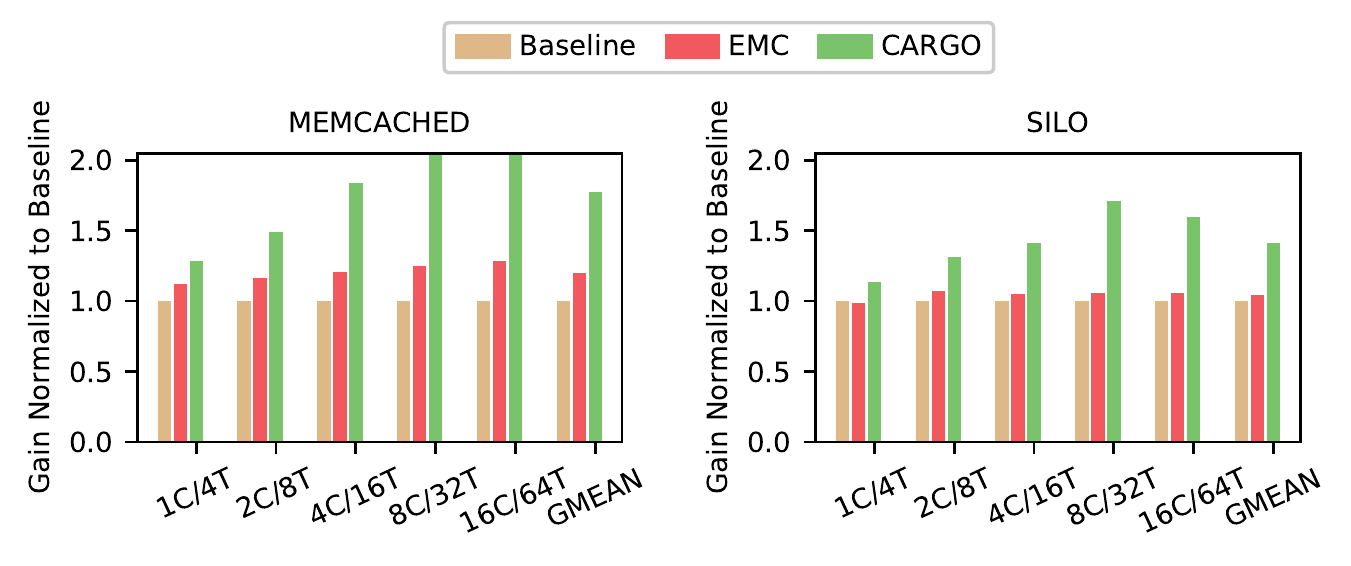}}
		\caption{Power efficiency normalized to baseline for 1 to 16 core configuration} 
		\label{eval-power-efficiency}
		\vspace{-20pt}
	\end{figure}		
	Since, our proposal introduces small area and power overhead (discussed in subsection~\ref{overhead}); improvement in throughput for CARGO directly translates to improved power-efficiency. Figure~\ref{eval-power-efficiency} shows normalized power-efficiency for CARGO and EMC. On average, CARGO achieves 2.9X and 2.2X improvement for memcached and Silo, respectively. Since, EMC falls behind CARGO in terms of throughput, it also lags in power-efficiency improvement. Across the board, EMC achieves 8\% improvement for Silo. We found that EMC fails to improve Silo due to its heavy use of software prefetching. Since, EMC triggers execution of a dependent load instruction only when a blocking load reaches the head of the ROB, a non-blocking load, such as, S/W prefetch delays dependence chain traversal and hides only a part of overall load latency. Because of this reason, EMC is unable to achieve expected performance. In contrast to EMC, CARGO identifies entire code region needed to generate loads that frequently miss in the cache hierarchy and executes them early from the NIC. As a result, CARGO achieves superior coverage than EMC and leads to much higher reduction in request processing latency.  	
	\subsection{Component-wise Analysis}
	\label{eval-perf-comp}
	\begin{figure}[!ht]
		\centering
		\fcolorbox{black}{white}{\includegraphics[width=.4\textwidth]{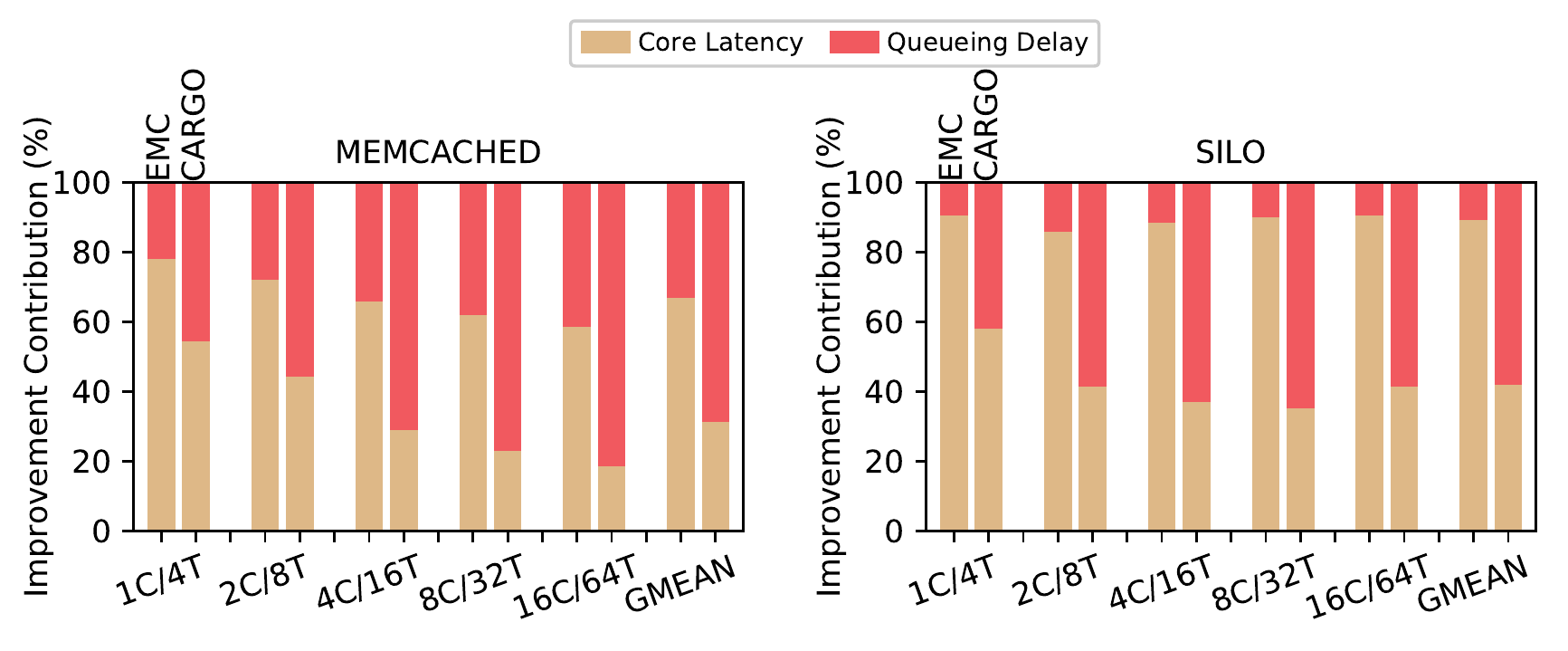}}
		\caption{Percent of improvement contributed by core and queuing delay.}
		\label{eval-core-queue-delay}
	        \vspace{-10pt}	
	\end{figure}
\begin{figure}[!ht]
		\centering
		\fcolorbox{black}{white}{\includegraphics[width=.4\textwidth]{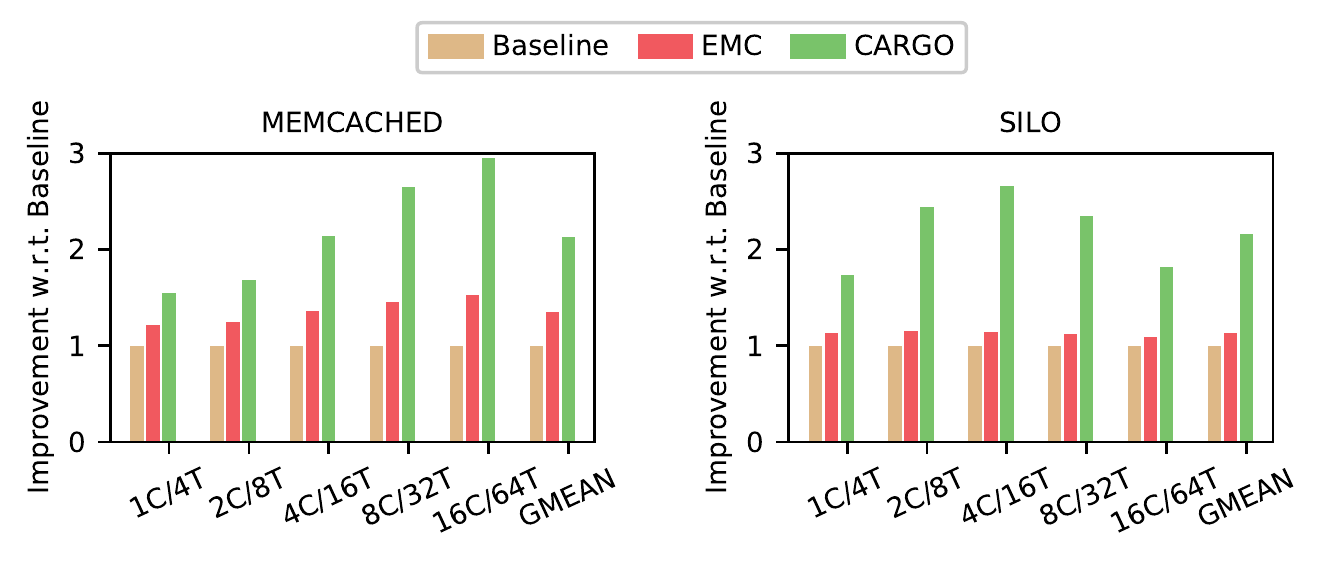}}
		\caption{DRAM bandwidth utilization improvement normalized to baseline for 1 to 16 core configuration}
		\vspace{-10pt}
		\label{eval-bw-utilization}
		\vspace{-10pt}
	\end{figure}
\begin{figure}[!ht]
		\centering
		\fcolorbox{black}{white}{\includegraphics[width=.4\textwidth]{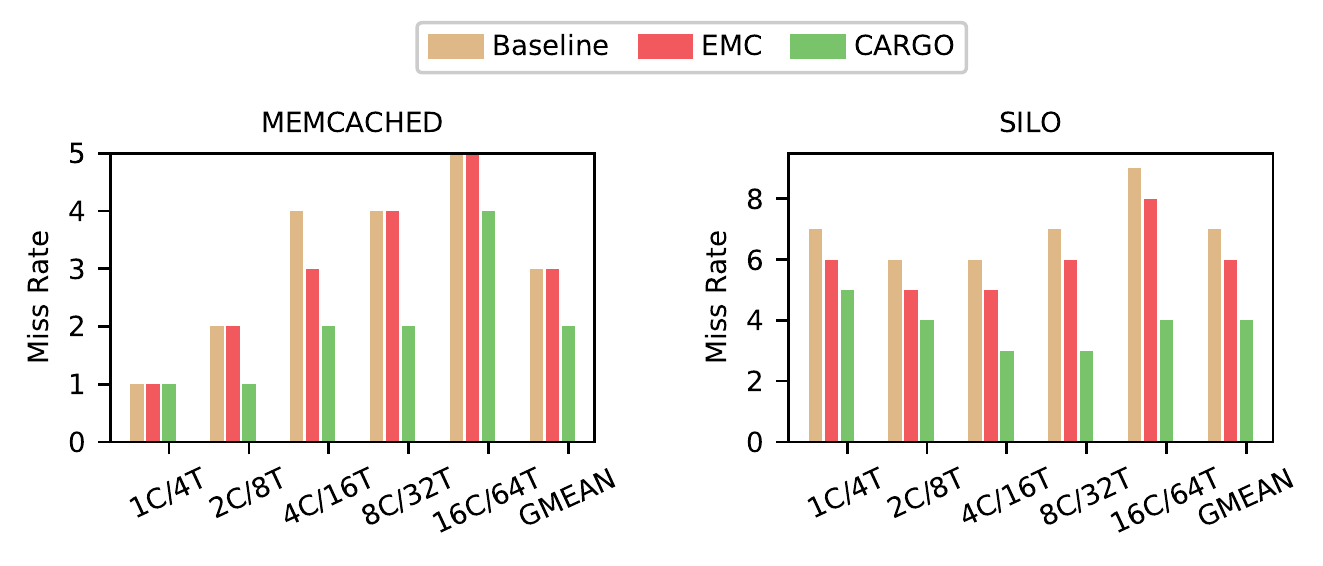}}
		\caption{Reduction in L1 miss rate for 1 to 16 core configuration}
		\label{eval-l1-miss}
		\vspace{-15pt}
	\end{figure}

	\noindent
	In this subsection we provide a more detailed analysis of performance improvement achieved by EMC and CARGO. We divide overall improvement into queuing and core components separately. Figure~\ref{eval-core-queue-delay} shows the percentage contribution of core and queuing delay for EMC and CARGO. On average, with EMC, queuing delay contributes 34\% and 11\% along with 66\% and 89\% contribution coming from core for memcached and Silo, respectively. Whereas, CARGO achieves 68\%, 58\% improvement from queuing delay and 32\%, 42\% improvement from core for memcached and Silo, respectively.	
	\begin{figure}[!ht]
		\centering
		\fcolorbox{black}{white}{\includegraphics[width=.4\textwidth]{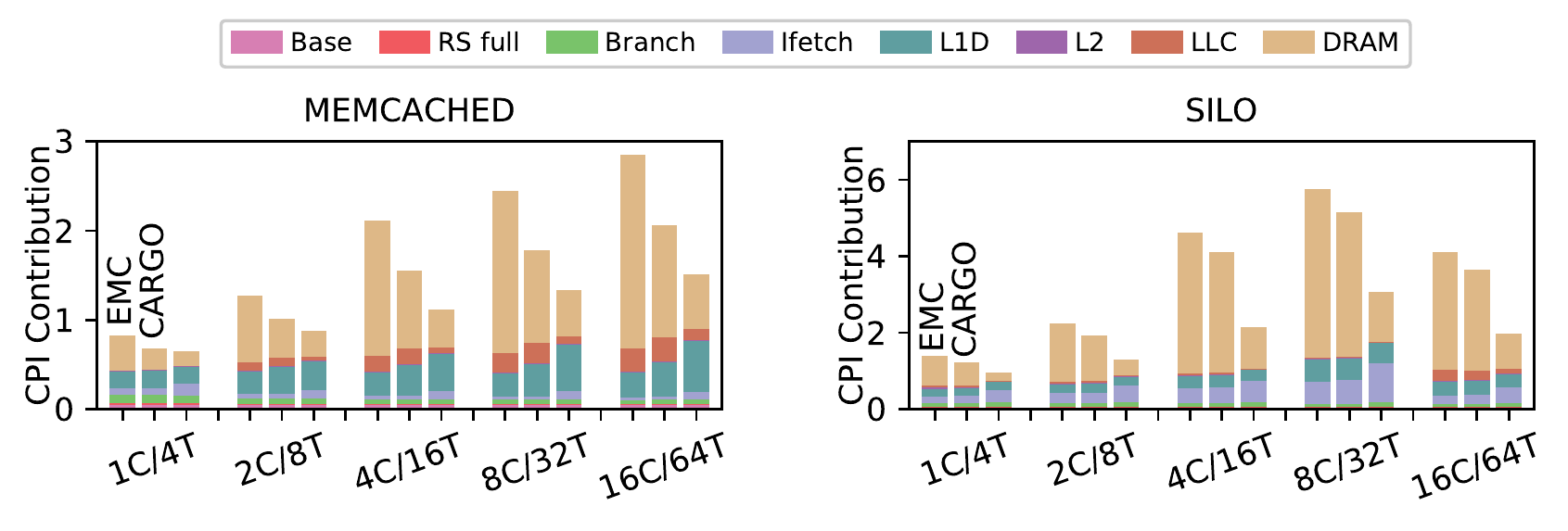}}
		\caption{CPI stack for 1 to 16 core configuration}
		\label{eval-cpi-stack}
	        \vspace{-10pt}
	\end{figure}	
	Core latency improvement in CARGO comes from both improved cache hit-rate and higher DRAM bandwidth utilization. To quantify the gains contributed by different core components,~Figure~\ref{eval-cpi-stack} presents the CPI contribution of different core components for baseline, EMC and CARGO. As the figure shows, across the board, DRAM and cache access contribute highest number of cycles. Moreover, as performance improves contribution of DRAM component goes down. Figure~\ref{eval-bw-utilization} shows the improvement in DRAM bandwidth utilization for different configurations. Compared to the baseline, CARGO improves the DRAM bandwidth utilization by 2.13X and 2.17X for memcached and Silo, respectively. However, we note that in some of the cases (4C/16T, 8C/32T, and 16C/64T configuration for memcached) filling data in L1 cache results in higher CPI contribution despite improved L1 hit-rate. We found that with higher L1 hit-rate, increased number of instructions find data in the L1 cache and thus CPI contributed by L1 access increases.To confirm this observation, Figure~\ref{eval-l1-miss} presents the miss-rate observed for baseline, EMC, and CARGO for different configurations. On average, CARGO reduces L1 miss rate by 1\% and 3\% for memcached and Silo, respectively. Thus, the increased CPI doesn't result from higher L1 miss rate but from higher number of instruction finding data in L1 cache.  

	
%
      
	\subsection{Execution, Area, Power, and PCIe Bandwidth Overhead:} 	
	\noindent
	Executing user routine and the critical region (presented in Section~\ref{cargo}) for each incoming packet introduces instruction overhead in packet processing. However, since, user routine is a small part of the application executing at the core, overhead incurred is very small. On average, across all workloads, user regions execute 56 additional instructions/packet, which is 4\% of the total instructions executed during complete packet processing at the NIC. Compared to a user routine, overhead incurred by the critical region is smaller. On average, 26 instructions with 5 memory operations are executed for each packet.
        \label{overhead}
	  \begin{figure}[!ht]
	    \centering
	    \fcolorbox{black}{white}{\includegraphics[width=.4\textwidth]{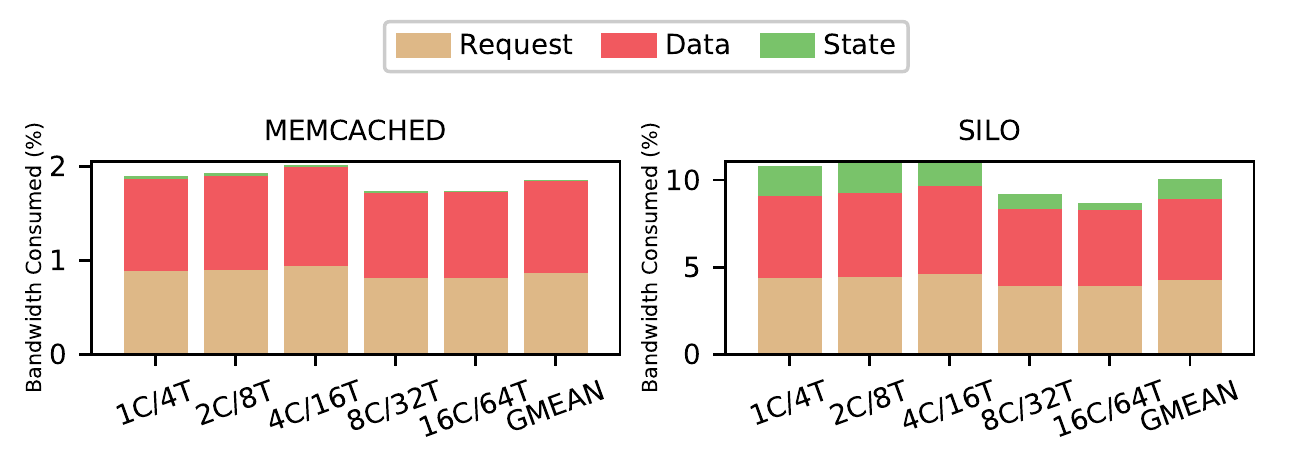}}
	    \caption{PCIe bandwidth overhead for 1 to 16 core configuration}
	    \label{eval-pcie-bw}
	    \vspace{-10pt}
	 \end{figure}		
	The area overhead in our proposal comes from the structures used to identify critical instructions and register values. The critical instruction cache is organized as a 16-way set-associative structure, where each entry takes 13 bytes of storage (8 byte PC, two 4 bit predecessor instructions, a ready bit, a valid bit and access counter for choosing the replacement candidate). To keep track of predecessor instructions, we use a 16 entry Register-to-PC map with one entry per architectural register. These two structures together consume 3.5 KB. The structure used for register state and register value predictor consume 272 bytes and 3 KB (48 entry level 1 (20 byte / entry) and 132 entry level 2 (16 byte / entry)), respectively. Overall, storage overhead of our proposal is slightly above 7 KB and consumes .4 W of additional power.
	
	Transferring and executing critical region from NIC also consumes PCIe bandwidth. Figure~\ref{eval-pcie-bw} quantifies the percent PCIe bandwidth consumed due to transfer of critical region and associated register state (State), issue of memory requests (Request), and the associated data transfer (Data). On average, across all operations, CARGO consumes 4.5\% bandwidth. Among all components transferring state incurs smallest overhead (less than 1\%). Whereas, Request and Data components contribute equally (1.9\% and 2.1\% for memcached and silo, respectively).    
	\subsection{Security Implication}
	\label{security}
	\noindent Since, our proposal doesn't transmit fetched data back to a requester, it doesn't leak any information. 
	But, like any DMA capable CPIe device, it can be used to launch side-channel attacks. However, Such attacks can be mitigated by strict IOMMU policies that 
	restrict access to a memory region by an I/O device. Apart from that, injecting data into the core cache also introduces risk of DoS attacks. However, since, datacenter 
	resources are usually partitioned to enforce QoS guarantees. Such mechanism will also defend our proposal from such attacks. In summary, CARGO doesn't introduce 
	any new security threat. However, it does suffer from side-channel and DoS attacks possible from PCIe devices in an shared environment of a datacenter and can be 
	mitigated using existing mechanisms. 
	\section{Conclusion}
	\label{conclusion}
	\noindent
	Network bound datacenter workloads not only suffer from NIC and interrupt processing inefficiency but also show poor ILP and MLP characteristics due to their pointer intensive design. Our study shows that compared to an FPGA based accelerator, a high-performance CPU with an efficient on-chip cache hierarchy can achieve much higher power-efficiency. However, long memory access latency, due to irregular access patterns, incurs long queuing latency and eventually degrades both throughput and the power efficiency of these applications. 
	
	To overcome these limitations, in this work, we have presented a novel NIC-Core co-design. Our proposal judiciously identifies critical instructions and the register state causing long latency memory accesses. Periodically, these instructions, along with register state, are sent to the NIC. The NIC executes the critical region for each incoming packet and prefetches data into L1 cache. With a modest amount of storage overhead, our proposal improves latency, throughput, and  energy of an in-memory key-value and a transactional database store by 2.7X, 2.7X, and 1.5X, respectively.    
	

\begin{thebibliography}{9}
\bibitem{swimp}	
Sam Ainsworth and Timothy M. Jones. 2017. Software prefetching for indirect memory accesses. In Proceedings of the 2017 International Symposium on Code Generation and Optimization (CGO '17). IEEE Press, Piscataway, NJ, USA, 305-317.

\bibitem{ix}
Adam Belay, George Prekas, Mia Primorac, Ana Klimovic, Samuel Grossman, Christos Kozyrakis, and Edouard Bugnion. 2016. The IX Operating System: Combining Low Latency, High Throughput, and Efficiency in a Protected Dataplane. ACM Trans. Comput. Syst. 34, 4, Article 11 (December 2016), 39 pages.

\bibitem{dataflowmemcached}
M. Blott, K. Karras, L. Liu, K. Vissers, J. Bär, and Z. István, "Achieving 10Gbps line-rate key-value stores with FPGAs," in HotCloud, 2013.

\bibitem{sniper}
Trevor E. Carlson, Wim Heirman, Lieven Eeckhout. Sniper: Exploring the Level of Abstraction for Scalable and Accurate Parallel Multi-Core Simulation. International Conference for High Performance Computing, Networking, Storage and Analysis (SC).

\bibitem{snipertaco}
Trevor E. Carlson, Wim Heirman, Stijn Eyerman, Ibrahim Hur, and Lieven Eeckhout. 2014. An Evaluation of High-Level Mechanistic Core Models. ACM Trans. Archit. Code Optim. 11, 3, Article 28 (August 2014), 25 pages.

\bibitem{hwdatabase}
Jared Casper and Kunle Olukotun. 2014. Hardware acceleration of database operations. In Proceedings of the 2014 ACM/SIGDA international symposium on Field-programmable gate arrays (FPGA '14). ACM, New York, NY, USA, 151-160.

\bibitem{hashjoinprefetching}
Shimin Chen, Anastassia Ailamaki, Phillip B. Gibbons, and Todd C. Mowry. 2007. Improving hash join performance through prefetching. ACM Trans. Database Syst. 32, 3, Article 17 (August 2007).

\bibitem{amdahls-law-tail-latency}
Christina Delimitrou and Christos Kozyrakis. 2018. Amdahl's law for tail latency. Commun. ACM 61, 8 (July 2018), 65-72.

\bibitem{clearingthecloud}
Michael Ferdman, Almutaz Adileh, Onur Kocberber, Stavros Volos, Mohammad Alisafaee, Djordje Jevdjic, Cansu Kaynak, Adrian Daniel Popescu, Anastasia Ailamaki, and Babak Falsafi. 2012. Clearing the clouds: a study of emerging scale-out workloads on modern hardware. In Proceedings of the seventeenth international conference on Architectural Support for Programming Languages and Operating Systems (ASPLOS XVII). ACM, New York, NY, USA, 37-48. 

\bibitem{fields}
Brian Fields, Shai Rubin, and Rastislav Bodík. 2001. Focusing processor policies via critical-path prediction. In Proceedings of the 28th annual international symposium on Computer architecture (ISCA '01). ACM, New York, NY, USA, 74-85.

\bibitem{steeringpatent}
Stephen D. Glaser, Mark D. Hummel, Methods and apparatus for injecting pci express traffic into host cache memory using a bit mask in the transaction layer steering tag. United States  Patent US 2013/0173834 A1, Publication Date: July 4, 2013.

\bibitem{threadvscache}
Zvika Guz, Oved Itzhak, Idit Keidar, Avinoam Kolod, Avi Mendelson, and Uri C. Weiser. Threads vs. caches: modeling the behavior of parallel workloads. In International Conference on Computer Design, October 2010. 

\bibitem{qmodel}
M. Harchol-Balter,Performance Modeling and Design of Computer Systems: Queueing Theory in Action, 2013

\bibitem{emc}
Milad Hashemi, Khubaib, Eiman Ebrahimi, Onur Mutlu, and Yale N. Patt. 2016. Accelerating dependent cache misses with an enhanced memory controller. In Proceedings of the 43rd International Symposium on Computer Architecture (ISCA '16). IEEE Press, Piscataway, NJ, USA, 444-455.

\bibitem{cre}
Milad Hashemi, Onur Mutlu, and Yale N. Patt. 2016. Continuous runahead: transparent hardware acceleration for memory intensive workloads. In The 49th Annual IEEE/ACM International Symposium on Microarchitecture (MICRO-49). IEEE Press, Piscataway, NJ, USA, Article 61, 12 pages.

\bibitem{memcachedgpu}
Tayler H. Hetherington, Mike O'Connor, and Tor M. Aamodt. 2015. MemcachedGPU: scaling-up scale-out key-value stores. In Proceedings of the Sixth ACM Symposium on Cloud Computing (SoCC '15). ACM, New York, NY, USA, 43-57. 

\bibitem{linkedlistnearmemory}
Byungchul Hong, Gwangsun Kim, Jung Ho Ahn, Yongkee Kwon, Hongsik Kim, and John Kim. 2016. Accelerating Linked-list Traversal Through Near-Data Processing. In Proceedings of the 2016 International Conference on Parallel Architectures and Compilation (PACT '16). ACM, New York, NY, USA, 113-124.

\bibitem{dca}
Ram Huggahalli, Ravi Iyer, and Scott Tetrick. 2005. Direct Cache Access for High Bandwidth Network I/O. In Proceedings of the 32nd annual international symposium on Computer Architecture (ISCA '05). IEEE Computer Society, Washington, DC, USA, 50-59.

\bibitem{fpgahash}
Zsolt István, Gustavo Alonso, Michaela Blott, and Kees Vissers. 2015. A Hash Table for Line-Rate Data Processing. ACM Trans. Reconfigurable Technol. Syst. 8, 2, Article 13 (March 2015), 15 pages. 

\bibitem{ima}
Akanksha Jain and Calvin Lin. 2013. Linearizing irregular memory accesses for improved correlated prefetching. In Proceedings of the 46th Annual IEEE/ACM International Symposium on Microarchitecture (MICRO-46). ACM, New York, NY, USA, 247-259.

\bibitem{markov}
Doug Joseph and Dirk Grunwald. 1997. Prefetching using Markov predictors. In Proceedings of the 24th annual international symposium on Computer architecture (ISCA '97). ACM, New York, NY, USA, 252-263.

\bibitem{mobilesearch}
Vijay Janapa Reddi, Benjamin C. Lee, Trishul Chilimbi, and Kushagra Vaid. 2010. Web search using mobile cores: quantifying and mitigating the price of efficiency. SIGARCH Comput. Archit. News 38, 3 (June 2010), 314-325.

\bibitem{criticalityvslocality}
Roy Dz-ching Ju, Alvin R. Lebeck, Chris Wilkerson. 2001. Locality vs. criticality. In Proceedings of the 28th annual international symposium on Computer architecture (ISCA '01). ACM, New York, NY, USA, 132-143.

\bibitem{warehousescale}
Svilen Kanev, Juan Pablo Darago, Kim Hazelwood, Parthasarathy Ranganathan, Tipp Moseley, Gu-Yeon Wei, and David Brooks. 2015. Profiling a warehouse-scale computer. In Proceedings of the 42nd Annual International Symposium on Computer Architecture (ISCA '15). ACM, New York, NY, USA, 158-169. 

\bibitem{meetthewalker}
Onur Kocberber, Boris Grot, Javier Picorel, Babak Falsafi, Kevin Lim, and Parthasarathy Ranganathan. 2013. Meet the walkers: accelerating index traversals for in-memory databases. In Proceedings of the 46th Annual IEEE/ACM International Symposium on Microarchitecture (MICRO-46). ACM, New York, NY, USA, 468-479. 

\bibitem{flexnic}
Antoine Kaufmann, SImon Peter, Naveen Kr. Sharma, Thomas Anderson, and Arvind Krishnamurthy. 2016. High Performance Packet Processing with FlexNIC. In Proceedings of the Twenty-First International Conference on Architectural Support for Programming Languages and Operating Systems (ASPLOS '16). ACM, New York, NY, USA, 67-81.

\bibitem{niclatency}
Steen Larsen, Parthasarathy Sarangam, Ram Huggahalli, and Siddharth Kulkarni. 2009. Architectural breakdown of end-to-end latency in a TCP/IP network. Int. J. Parallel Program. 37, 6 (December 2009), 556-571.

\bibitem{highserverutil}
Jacob Leverich and Christos Kozyrakis. 2014. Reconciling high server utilization and sub-millisecond quality-of-service. In Proceedings of the Ninth European Conference on Computer Systems (EuroSys '14). ACM, New York, NY, USA, Article 4, 14 pages.

\bibitem{kvdirect}
Bojie Li, Zhenyuan Ruan, Wencong Xiao, Yuanwei Lu, Yongqiang Xiong, Andrew Putnam, Enhong Chen, and Lintao Zhang. 2017. KV-Direct: High-Performance In-Memory Key-Value Store with Programmable NIC. In Proceedings of the 26th Symposium on Operating Systems Principles (SOSP '17). ACM, New York, NY, USA, 137-152.

\bibitem{billion}
Sheng Li, Hyeontaek Lim, Victor W. Lee, Jung Ho Ahn, Anuj Kalia, Michael Kaminsky, David G. Andersen, O. Seongil, Sukhan Lee, and Pradeep Dubey. 2015. Architecting to achieve a billion requests per second throughput on a single key-value store server platform. SIGARCH Comput. Archit. News 43, 3 (June 2015), 476-488.

\bibitem{tailsoftail}
Jialin Li, Naveen Kr. Sharma, Dan R. K. Ports, and Steven D. Gribble. 2014. Tales of the Tail: Hardware, OS, and Application-level Sources of Tail Latency. In Proceedings of the ACM Symposium on Cloud Computing (SOCC '14). ACM, New York, NY, USA, , Article 9 , 14 pages. 

\bibitem{agressivepipeline}
Zhaoshi Li, Leibo Liu, Yangdong Deng, Shouyi Yin, Yao Wang, and Shaojun Wei. 2017. Aggressive Pipelining of Irregular Applications on Reconfigurable Hardware. SIGARCH Comput. Archit. News 45, 2 (June 2017), 575-586.

\bibitem{mica}
Hyeontaek Lim, Dongsu Han, David G. Andersen, and Michael Kaminsky. 2014. MICA: a holistic approach to fast in-memory key-value storage. In Proceedings of the 11th USENIX Conference on Networked Systems Design and Implementation (NSDI'14). USENIX Association, Berkeley, CA, USA, 429-444.

\bibitem{thinserversmartpipes}
Kevin Lim, David Meisner, Ali G. Saidi, Parthasarathy Ranganathan, and Thomas F. Wenisch. 2013. Thin servers with smart pipes: designing SoC accelerators for memcached. SIGARCH Comput. Archit. News 41, 3 (June 2013), 36-47.

\bibitem{incbricks}
Ming Liu, Liang Luo, Jacob Nelson, Luis Ceze, Arvind Krishnamurthy, and Kishore Atreya. 2017. IncBricks: Toward In-Network Computation with an In-Network Cache. SIGARCH Comput. Archit. News 45, 1 (April 2017), 795-809.

\bibitem{compprefrecursive}
hi-Keung Luk and Todd C. Mowry. 1996. Compiler-based prefetching for recursive data structures. In Proceedings of the seventh international conference on Architectural support for programming languages and operating systems (ASPLOS VII). ACM, New York, NY, USA, 222-233.

\bibitem{low-power-vs-energy-efficiency}
David Meisner and Thomas F. Wenisch. 2011. Does low-power design imply energy efficiency for data centers?. In Proceedings of the 17th IEEE/ACM international symposium on Low-power electronics and design (ISLPED '11). IEEE Press, Piscataway, NJ, USA, 109-114. 

\bibitem{runahead}
Onur Mutlu, Jared Stark, Chris Wilkerson, and Yale N. Patt. 2003. Runahead Execution: An Alternative to Very Large Instruction Windows for Out-of-Order Processors. In Proceedings of the 9th International Symposium on High-Performance Computer Architecture (HPCA '03). IEEE Computer Society, Washington, DC, USA, 129-. 

\bibitem{pcie}
Rolf Neugebauer, Gianni Antichi, José Fernando Zazo, Yury Audzevich, Sergio López-Buedo, and Andrew W. Moore. 2018. Understanding PCIe performance for end host networking. In Proceedings of the 2018 Conference of the ACM Special Interest Group on Data Communication (SIGCOMM '18). ACM, New York, NY, USA, 327-341. 

\bibitem{catch}
Anant Vithal Nori, Jayesh Gaur, Siddharth Rai, Sreenivas Subramoney, and Hong Wang. 2018. Criticality aware tiered cache hierarchy: a fundamental relook at multi-level cache hierarchies. In Proceedings of the 45th Annual International Symposium on Computer Architecture (ISCA '18). IEEE Press, Piscataway, NJ, USA, 96-109.

\bibitem{cloudsuit}
Palit, Tapti, Yongming Shen and Michael Ferdman. “Demystifying cloud benchmarking.” 2016 IEEE International Symposium on Performance Analysis of Systems and Software (ISPASS) (2016): 122-132.

\bibitem{zygos}
George Prekas, Marios Kogias, and Edouard Bugnion. 2017. ZygOS: Achieving Low Tail Latency for Microsecond-scale Networked Tasks. In Proceedings of the 26th Symposium on Operating Systems Principles (SOSP '17). ACM, New York, NY, USA, 325-341.

\bibitem{lowlatency}
Stephen M. Rumble, Diego Ongaro, Ryan Stutsman, Mendel Rosenblum, and John K. Ousterhout. 2011. It's time for low latency. In Proceedings of the 13th USENIX conference on Hot topics in operating systems (HotOS'13). USENIX Association, Berkeley, CA, USA, 11-11.

\bibitem{seng}
John S. Seng, Eric S. Tune, and Dean M. Tullsen. 2001. Reducing power with dynamic critical path information. In Proceedings of the 34th annual ACM/IEEE international symposium on Microarchitecture (MICRO 34). IEEE Computer Society, Washington, DC, USA, 114-123. 

\bibitem{panic}
Brent Stephens, Aditya Akella, Michael M. Swift. (2018). Your Programmable NIC Should be a Programmable Switch. 36-42. 

\bibitem{elasticflow}
Mingxing Tan, Gai Liu, Ritchie Zhao, Steve Dai, and Zhiru Zhang. 2015. ElasticFlow: A Complexity-Effective Approach for Pipelining Irregular Loop Nests. In Proceedings of the IEEE/ACM International Conference on Computer-Aided Design (ICCAD '15). IEEE Press, Piscataway, NJ, USA, 78-85. 

\bibitem{silo}
Stephen Tu, Wenting Zheng, Eddie Kohler, Barbara Liskov, and Samuel Madden. 2013. Speedy transactions in multicore in-memory databases. In Proceedings of the Twenty-Fourth ACM Symposium on Operating Systems Principles (SOSP '13). ACM, New York, NY, USA, 18-32. 

\bibitem{tune}
Eric Tune, Dean M. Tullsen, and Brad Calder. 2002. Quantifying Instruction Criticality. In Proceedings of the 2002 International Conference on Parallel Architectures and Compilation Techniques (PACT '02). IEEE Computer Society, Washington, DC, USA, 104-. 

\bibitem{rdbmsenergy}
Dimitris Tsirogiannis, Stavros Harizopoulos, and Mehul A. Shah. 2010. Analyzing the energy efficiency of a database server. In Proceedings of the 2010 ACM SIGMOD International Conference on Management of data (SIGMOD '10). ACM, New York, NY, USA, 231-242.

\bibitem{p4fpga}
Han Wang, Robert Soulé, Huynh Tu Dang, Ki Suh Lee, Vishal Shrivastav, Nate Foster, and Hakim Weatherspoon. 2017. P4FPGA: A Rapid Prototyping Framework for P4. In Proceedings of the Symposium on SDN Research (SOSR '17). ACM, New York, NY, USA, 122-135. 

\bibitem{pointeronfpga}
Gabriel Weisz, Joseph Melber, Yu Wang, Kermin Fleming, Eriko Nurvitadhi, and James C. Hoe. 2016. A Study of Pointer-Chasing Performance on Shared-Memory Processor-FPGA Systems. In Proceedings of the 2016 ACM/SIGDA International Symposium on Field-Programmable Gate Arrays (FPGA '16). ACM, New York, NY, USA, 264-273.

\bibitem{10gignic}
Paul Willman, Hyong-youb Kim, Scott Rixner, V. S. Pai (2005). An Efficient Programmable 10 Gigabit Ethernet Network Interface Card. 96- 107. 

\bibitem{memcachdchar}
Y. Xu, E. Frachtenberg, S. Jiang and M. Paleczny, "Characterizing Facebook's Memcached Workload," in IEEE Internet Computing, vol. 18, no. 2, pp. 41-49, Mar.-Apr. 2014.

\bibitem{bubbleflux}
Hailong Yang, Alex Breslow, Jason Mars, and Lingjia Tang. 2013. Bubble-flux: precise online QoS management for increased utilization in warehouse scale computers. SIGARCH Comput. Archit. News 41, 3 (June 2013), 607-618.

\bibitem{stackmap}
Kenichi Yasukata, Michio Honda, Douglas Santry, and Lars Eggert. 2016. StackMap: low-latency networking with the OS stack and dedicated NICs. In Proceedings of the 2016 USENIX Conference on Usenix Annual Technical Conference (USENIX ATC '16). USENIX Association, Berkeley, CA, USA, 43-56.

\bibitem{imp}
Xiangyao Yu, Christopher J. Hughes, Nadathur Satish, and Srinivas Devadas. 2015. IMP: indirect memory prefetcher. In Proceedings of the 48th International Symposium on Microarchitecture (MICRO-48). ACM, New York, NY, USA, 178-190.

\bibitem{minnow}
Dan Zhang, Xiaoyu Ma, Michael Thomson, and Derek Chiou. 2018. Minnow: Lightweight Offload Engines for Worklist Management and Worklist-Directed Prefetching. In Proceedings of the Twenty-Third International Conference on Architectural Support for Programming Languages and Operating Systems (ASPLOS '18). ACM, New York, NY, USA, 593-607.

\bibitem{memoization}
Guowei Zhang and Daniel Sanchez. 2019. Leveraging Caches to Accelerate Hash Tables and Memoization. In Proceedings of the 52nd Annual IEEE/ACM International Symposium on Microarchitecture (MICRO '52). ACM, New York, NY, USA, 440-452. 

\bibitem{megakv}
Kai Zhang, Kaibo Wang, Yuan Yuan, Lei Guo, Rubao Lee, and Xiaodong Zhang. 2015. Mega-KV: a case for GPUs to maximize the throughput of in-memory key-value stores. Proc. VLDB Endow. 8, 11 (July 2015), 1226-1237. 

\bibitem{silosrc}
https://github.com/stephentu/silo

\bibitem{rfs}
https://www.kernel.org/doc/html/latest/networking/scaling.html


\end{thebibliography}


\end{document}